%% file: andersonMBL_arXiv_v1.tex
\documentclass[pra,aps,twocolumn,showpacs,superscriptaddress,footnoteinbib]{revtex4-1}

\usepackage{graphicx}
\usepackage{dcolumn}
\usepackage{bm}
\usepackage{amssymb,amsmath}
\usepackage{sidecap}
\usepackage{wrapfig}
\usepackage{natbib}

\usepackage{graphicx}
\usepackage{leftidx}
\usepackage{color}

\usepackage{bbold} 

\usepackage{wasysym} 

\usepackage{stackrel}

\usepackage{soul,xcolor}
\setstcolor{red}

\newcommand{\be}{\begin{equation}}
\newcommand{\ee}{\end{equation}}
\newcommand{\ba}{\begin{align}}
\newcommand{\ea}{\end{align}}
\newcommand{\sysb}{\left\{\begin{array}}
\newcommand{\syse}{\end{array}\right.}
\newcommand{\baa}{\begin{array}}
\newcommand{\eaa}{\end{array}}
\newcommand{\bs}{\begin{split}}
\newcommand{\es}{\end{split}}

\newcommand{\matb}{\left(\begin{array}}
\newcommand{\mate}{\end{array}\right)}

\newcommand{\rmd}{{\rm{d}}}

\newcommand{\rme}[1]{{\rm{e}}^{#1}}

\newcommand{\ha}{\frac{1}{2}}

\newcommand{\lt}{\left(}
\newcommand{\rt}{\right)}
\newcommand{\lqq}{\left[}
\newcommand{\rqq}{\right]}
\newcommand{\lan}{\left\langle}
\newcommand{\ran}{\right\rangle}
\newcommand{\abs}[1]{\left| #1 \right|}

\newcommand{\av}[1]{\lan #1 \ran}
\newcommand{\norm}[1]{\left\| #1 \right\|}

\newcommand{\ket}[1]{\left| #1 \ran}
\newcommand{\bra}[1]{\lan #1 \right|}
\newcommand{\bracket}[2]{\lan #1 \right| \!\left. #2 \ran}
\newcommand{\proj}[1]{\ket{#1} \bra{#1}}

\newcommand{\nol}{\nonumber \\}

\newcommand{\prodl}[2]{\prod\limits_{#1}^{#2}}

\newcommand{\dar}{\downarrow}
\newcommand{\uar}{\uparrow}

\newcommand{\ind}{b}
\newcommand{\indd}{c}

\usepackage{amsthm}

\begin{document}

\title{Localization phenomena in interacting Rydberg lattice gases with position disorder}


\author{Matteo Marcuzzi}
\affiliation{School of Physics and Astronomy, University of Nottingham, Nottingham, NG7 2RD, United Kingdom}

\author{Ji{\v{r}}\'i Min\'a{\v r}}
\affiliation{School of Physics and Astronomy, University of Nottingham, Nottingham, NG7 2RD, United Kingdom}

\author{Daniel Barredo}
\affiliation{Laboratoire Charles Fabry, Institut d'Optique Graduate School, CNRS, Universit\'e Paris-Saclay, 91127 Palaiseau cedex, France}

\author{Sylvain de L{\'e}s{\'e}leuc}
\affiliation{Laboratoire Charles Fabry, Institut d'Optique Graduate School, CNRS, Universit\'e Paris-Saclay, 91127 Palaiseau cedex, France}

\author{Henning Labuhn}
\affiliation{Laboratoire Charles Fabry, Institut d'Optique Graduate School, CNRS, Universit\'e Paris-Saclay, 91127 Palaiseau cedex, France}

\author{Thierry Lahaye}
\affiliation{Laboratoire Charles Fabry, Institut d'Optique Graduate School, CNRS, Universit\'e Paris-Saclay, 91127 Palaiseau cedex, France}

\author{Antoine Browaeys}
\affiliation{Laboratoire Charles Fabry, Institut d'Optique Graduate School, CNRS, Universit\'e Paris-Saclay, 91127 Palaiseau cedex, France}

\author{Emanuele Levi}
\affiliation{School of Physics and Astronomy, University of Nottingham, Nottingham, NG7 2RD, United Kingdom}

\author{Igor Lesanovsky}
\affiliation{School of Physics and Astronomy, University of Nottingham, Nottingham, NG7 2RD, United Kingdom}

%
%

%
%

%
%


\begin{abstract}
Disordered systems provide paradigmatic instances of ergodicity breaking and localization phenomena. Here we explore the dynamics of excitations in a system of Rydberg atoms held in optical tweezers. The finite temperature produces an intrinsic uncertainty in the atomic positions, which translates into quenched correlated disorder in the interatomic interaction strengths. 
In a simple approach, the dynamics in the many-body Hilbert space can be understood in terms of a one-dimensional Anderson-like model with disorder on every other site, featuring both localized and delocalized states. We conduct an experiment on an eight-atom chain and observe a clear suppression of excitation transfer. Our experiment accesses a regime which is described by a two-dimensional Anderson model on a ``trimmed'' square lattice. Our results thus provide a concrete example in which the absence of excitation propagation in a many-body system is directly related to Anderson-like localization in the Hilbert space, which is believed to be the mechanism underlying many-body localization.
\end{abstract}

\maketitle


\section{Introduction}
\label{sec:Introduction}

In his seminal work Anderson showed \cite{Anderson1958} that the spectrum of a free electron subject to a sufficiently strongly disordered potential consists solely of spatially localized wavefunctions, a phenonemon subsequently coined Anderson localization. In one dimension, all states are localized even for arbitrarily small disorder, which prevents any charge transport \cite{Mott1961,Ishii1973}. Anderson localization has been now observed experimentally in a number of physical systems, such as electron gases \cite{Cutler:1969}, cold atoms in a speckle potential both in one \cite{Billy:2008,Roati:2008} and three \cite{Semeghini:2015} dimensions, thin film topological insulators \cite{Liao:2015} or molecular rotors \cite{Bitter:2016}.

An ongoing problem is the extension of the Anderson paradigm to many-body systems \cite{Altshuler1997,Gornyi2005,Basko2006,Oganesyan2007,Pal2010} including systems with long-range interactions \cite{Yao2014,Hauke2014,Burin2015,Smith2015}. In \cite{Gornyi2005,Basko2006} it is argued that for weakly-interacting electrons there is a temperature-driven metal-to-insulator transition, which can be interpreted as Anderson-like localization of many-body wave functions in the Fock basis. The localization of these wavefunctions then becomes a crucial element in understanding phenomena like ergodicity breaking and the emergence of so-called many-body localized phases. Here, contrary to the central assumption of statistical mechanics, a many-body system retains memory of its initial conditions even at long times \cite{Gogolin2011,Schreiber2015,Hauke2014}. Only very recently experiments have started to probe this physics in systems of cold fermions \cite{Schreiber2015} and ions \cite{Smith2015}.

In this work we employ Rydberg atoms in a chain of optical tweezers to explore a many-body system whose dynamical properties are governed by Anderson localization in Fock space, much like the mechanism envisioned for weakly interacting electron gases in Ref.~\cite{Basko2006}.
Remarkably, a connection arises between the Rydberg system and a one- or two-dimensional variant of the Anderson model. These models feature correlated and site-dependent disorder, the origin of which lies in the intrinsic uncertainty of the atomic positions within the tweezers. The spectrum of the generalized Anderson models includes localized as well as delocalized many-body wave functions on the Fock basis. In the one-dimensional case localization in Fock space translates into localization in real space; for the 2D case this is not necessarilty true, and a richer structure emerges.
We study experimentally the resulting suppression of excitation transfer in an elementary example of two atoms as well as in a chain of eight atoms.

\section{Experimental setup and model}
\label{sec:Setup}

\begin{figure}
  \includegraphics[width=\columnwidth]{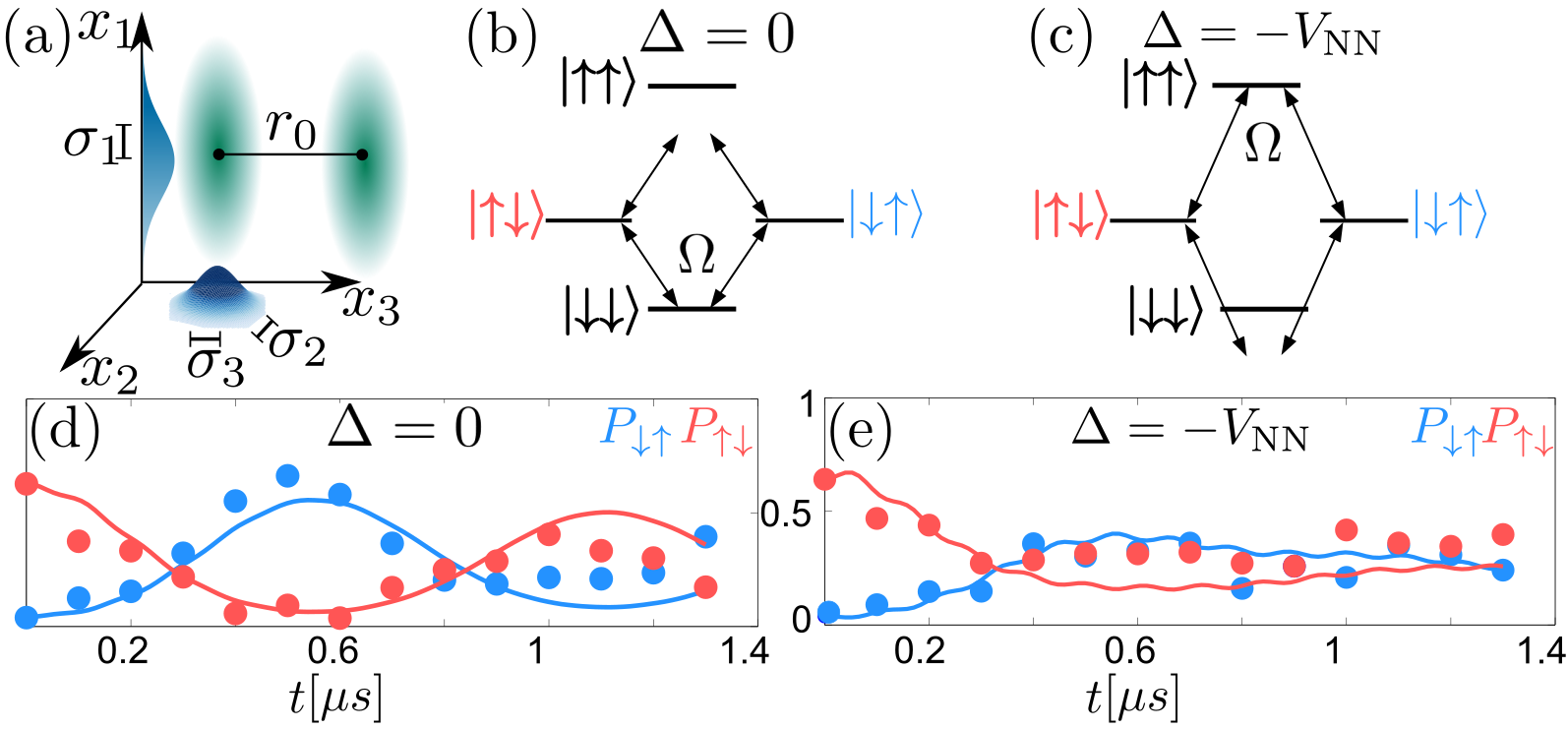}
  \caption{Two-atom setting. In panel (a) we show the setting of the system under consideration: The harmonic traps are ordered in a line along $x_3$ with average separation $r_0$ and widths $\sigma_i$ along the orthogonal directions $i=1,2,3$.
  Panels (b,c) show the level structure for the two-atom case for the resonant ($\Delta=0$) and facilitated ($\Delta = -V_{\rm NN}$) conditions respectively. The experimental data for the time evolution of the excitation probabilities $P_{\uparrow \downarrow}$, $P_{\downarrow \uparrow}$ are shown as full circles in panels (d,e) for the resonant and facilitated conditions respectively. The data are averaged over at least 100 realizations of the disorder, and the standard errors are smaller than the data points.
  The solid lines show the analytical solutions obtained by numerically solving the dynamics and averaging over 30 realizations of the disorder. The experiment is performed on two traps at distance $r_0=14.2\;\mu$m, each loaded with a single $^{87}$Rb atom. The measured atom temperature $T=50\;\mu$K, $\sigma_1=1\;\mu$m and $\sigma_{2,3}=120$ nm. The internal levels are $\ket{\downarrow}=\ket{5S_{1/2},F=2,M=2}$ and $\ket{\uparrow}=\ket{100\;D_{3/2},F=3,M=3}$ with $C_6 = - 2\pi \times 7.3 \times 10^7\;{\rm MHz}\,\mu{\rm m}^6$. Consequently, $V_{\rm NN} = C_6 / r_0^6 = -2 \pi \times 8.9 \,{\rm MHz}$, while the typical energy displacement is $\abs{\delta V} \sim 2\pi \times 0.64 \,{\rm MHz}$. The Rabi frequency of the driving laser is $\Omega = 2 \pi \times 1.25$ MHz.
  }
\label{fig:1}
\end{figure}

We consider a chain of tight optical traps, where each trap is loaded with a single atom \cite{Nogrette2014, Labuhn2014, Barredo2015, Labuhn2015}. 
In Fig.~\ref{fig:1}(a) we show an example of such a setting for two atoms. We label the Cartesian coordinates with an index $i = 1,2,3$ and fix them in such a way that the chain lies along direction $3$. The average separation between contiguous traps is $\mathbf{r}_0=(0,0,r_0)$.
We describe the Rydberg atoms as effective two-level systems \cite{Rydberg2} consisting of the electronic ground state $\ket{\dar}$ and a Rydberg excited state (or ``excitation'') $\ket{\uar}$. In the following, we shall refer to the product states of $\ket{\uar}$ and $\ket{\dar}$ spins as our ``Fock basis''.
The atoms are driven by laser light with Rabi frequency $\Omega$, and relative detuning $\Delta$.
A cartoon of a two-atom level structure is shown in Fig.~\ref{fig:1}(b,c).
The excitations mutually interact via a van-der-Waals potential $V(\abs{\mathbf{r}} ) = C_6/\left|\mathbf{r}\right|^6$ \cite{Beguin2013, Rydberg2}.
The Hamiltonian of the system, in a rotating wave approximation, reads
\be
	H =  \sum_k   \lqq  \frac{\Omega}{2} \sigma^x_k +   \Delta n_k   + \sum_{l>k} V(\abs{\mathbf{r}_k - \mathbf{r}_l} )   n_k n_l\rqq
	\label{eq:H_0}
\ee
where $\sigma^x_k = \ket{\uar_k} \bra{\dar_k} + \ket{\dar_k} \bra{\uar_k}$ and $n_k = \proj{\uar_k}$. 
Setting the origin in the center of the first trap, we can express the $k$-th atom position as $\mathbf{r}_k=(k-1)\mathbf{r}_0+\delta \mathbf{r}_k$.
The displacements $\delta\mathbf{r}_k$ originate from the finite temperature $T$ of the atoms and constitute an intrinsic source of randomness. If $T$ is sufficiently low, the atoms, which are frozen during the experiment, mostly occupy the harmonic part of the traps. Hence, their distribution is approximately a Gaussian with widths $\sigma_i = \sqrt{k_B T/(m\omega_i^2)}$ along the directions $i = 1,2,3$, where $m$ is the mass of a single atom and $\omega_i$ the trapping frequency (see Appendix \ref{app:distribution}).
The randomness thereby appears in equation \eqref{eq:H_0} via the interaction term, which depends on the random distances $d_{k,l} = \abs{\mathbf{r}_{k+l} - \mathbf{r}_k} = \abs{l\mathbf{r}_0 + \delta \mathbf{r}_{k+l} - \delta \mathbf{r}_{k}}$. For later purposes, we also introduce the energy displacements $\delta V_k \equiv  V(d_{k,1}) - V(r_0)$. Note that these differences are not independent: for instance, both $d_{k+1,1}$ and $d_{k,1}$ depend on $\mathbf{r}_{k+1}$, which generates correlation between them (we further address this issue in Appendix \ref{app:correlations}). 

\section{Two-atom case}
\label{sec:Two-atoms}

We start by illustrating the effect of the randomness in a two-atom setting. Considering first $\Delta=0$ (atomic level structure shown in Fig.~\ref{fig:1}(b)), the two atomic states $\ket{\uparrow \downarrow}, \ket{\downarrow \uparrow}$ are resonant with $\ket{\downarrow \downarrow}$, while the interaction brings $\ket{\uar \uar}$ off resonance and thus decouples it from the dynamics. Since the disorder only acts on $\ket{\uar \uar}$, a dynamics starting from $\ket{\dar \dar}$, $\ket{\uar \dar}$, $\ket{\dar \uar}$, or combinations thereof, is not affected by it.
In the experiment, after preparing the system in the $\ket{\uar \dar}$ state \cite{Labuhn2014}, the evolution resembles a coherent oscillation of the initial excitation between the two atoms. This is shown in Fig.~\ref{fig:1}(d), where we display the excitation probabilities $P_{\uparrow \downarrow} = \av{n_1 (1-n_2)}$, $P_{\downarrow \uparrow }  = \av{(1-n_1) n_2}$ as functions of time. The presence of the disorder becomes apparent instead when driving the system through the $\ket{\uparrow \uparrow}$ resonance. This is achieved by setting $\Delta=-V_{\mathrm{NN}}$, the so-called ``facilitation condition'' \cite{Ates07, Amthor2010, Lesanovsky14, PRL-KinC, Valado2015}, where $V_{\mathrm{NN}}=V(r_0)$ is the nearest-neighbor interaction energy in the absence of disorder, Fig. \ref{fig:1}(c). Here, the amplitude of the oscillations of $P_{\downarrow\uparrow}$ and $P_{\uparrow\downarrow}$ is clearly suppressed, Fig.~\ref{fig:1}(e). This means that the displacements $\delta \mathbf{r}_{1}$, $\delta \mathbf{r}_{2}$ are on average sufficiently large to bring the $\ket{\uparrow \uparrow}$ state off-resonance and in turn inhibit the propagation of the initial excitation (see Appendix \ref{app:correlations} for more details).

\begin{figure}
  \includegraphics[width=\columnwidth]{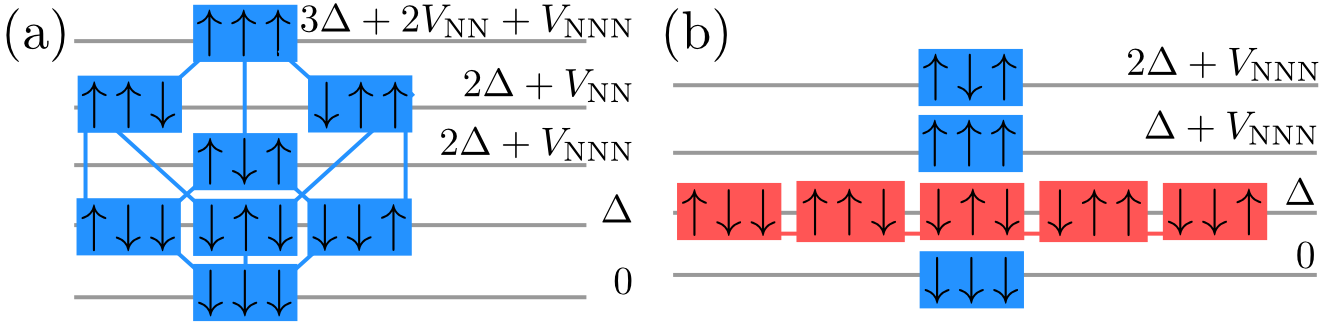}
  \caption{Fock space: structure and layering. Panels (a),(b) show the Fock space structure for three atoms respectively prior and after applying the facilitation condition. The states are ordered in rows corresponding to the associated energy. In panel (a) the internal structure of the Fock space is highlighted by linking states which can be connected by one spin flip. Panel (b) shows in red the resulting reduced Hilbert space in the regime we consider.
  }
\label{fig:2}
\end{figure}

\section{Generalization to many atoms}
\label{sec:Many-atoms}

In the following we will focus on the dynamics within a chain of atoms. To gain insight on the expected phenomena we will consider a simplified setting before turning to the actual experiment.
The Fock space for the model at hand can be depicted as a complex network of states.
This is sketched in Fig.~\ref{fig:2}(a) for three atoms: Only states which differ by a single spin flip are connected by Hamiltonian \eqref{eq:H_0} via the ``flipping'' ($\propto \Omega$) term.
Momentarily not accounting for the disorder, the states organize into energy layers, where we dub $V_{\rm NNN} = V(2r_0)$ the next-nearest-neighbor interactions and assume we can neglect all terms beyond this distance (i.e., we neglect $V(n r_0)$ for $n > 2$).

In the following we fix the facilitation condition $\Delta = -V_{\rm NN}$, which allows us to investigate the propagation of excitations in the presence of disorder. 
A remarkable simplification of the description ensues if we assume:
(i) \emph{large detuning} ($\Delta \gg \Omega$). This strongly suppresses unfacilitated transitions, i.e., spin flips not in the presence of a single excitation nearby.
(ii) \emph{strong next-nearest neighbor blockade} ($V(2r_0) \gg \Omega, \delta V_k$). Interactions at distance $2 r_0$ are supposed to be sufficiently strong to suppress the atomic transitions. In particular, we require this suppression to be much stronger than the one produced by the disorder. We also consider a tight confinement of the atoms, $\sigma_j \ll r_0$, such that, as in Fig.~\ref{fig:1}(e), the disorder can hinder, but not prevent transport entirely (i.e., $\delta V_k \lesssim \Omega$).

Under these conditions the states organize again in layers with large energy gaps approximately of the order of $V_{\rm NNN}$ or $ \Delta$. Within each layer, however, states are now separated by considerably smaller differences $\delta V_k$. We thereby neglect connections between different layers and retain only the intra-layer ones.
We sketch in Fig.~\ref{fig:2}(b) this layered structure for the network considered in Fig.~\ref{fig:2}(a).

We focus now on the highlighted (red) layer at energy $\Delta$, whose structure can be generalized in a straightforward manner to arbitrary chains with $L$ sites, as we show below.
We recall first that (i) implies that spins cannot be flipped if they do not have a \emph{single} excited neighbor. As a consequence, clusters of consecutive excitations can shrink or grow, but not merge or (dis)appear, i.e., the number $N_{\rm cl}$ of these clusters is conserved (see also the discussion in Appendix \ref{app:Hilbert}).
Condition (ii) implies instead that a spin next to two consecutive excitations cannot flip (e.g., $\ket{\uar \uar \dar} \leftrightarrow \ket{\uar \uar \uar}$ is forbidden); it then follows that the number $N_{\rm NNN}$ of excitation triples ($\uar \uar \uar$) is conserved.
The red layer in Fig.~\ref{fig:2}(b) corresponds to $N_{\rm cl} = 1$, $N_{\rm NNN} = 0$ as it exclusively includes states with a single excitation or a single pair of neighboring ones; in the following, the former kind will be denoted by odd integers, $\ket{2j - 1} \equiv \ket{ \dar_1 \ldots \dar_{j-1} \uar_j \dar_{j+1} \ldots \dar_{L}}$ ($j = 1 \ldots L$) whereas the latter by even integers, $\ket{2j} \equiv \ket{ \dar_1 \ldots \dar_{j-1} \uar_j \uar_{j+1} \dar_{j+2} \ldots \dar_{L}}$ ($j = 1 \ldots L-1$). The dynamics restricted to this layer can be described by an effective one-dimensional Anderson model \cite{Anderson1958}. In fact, the Hamiltonian connects these states sequentially ($\ldots \ket{2j-1} \leftrightarrow \ket{2j} \leftrightarrow \ket{2j+1} \ldots$), taking the form of a tight-binding model with sites labeled by $\ind = 1 \ldots 2L-1$ and a random potential $h_\ind = (1 + (-1)^\ind)\delta V_{\ind/2}/\Omega$ acting only on even ones. In this restricted space $H$ can be recast as (see Appendix \ref{app:Hilbert})
\be
H_{\mathrm{A}}=\frac{\Omega}{2}\sum_{\ind =1}^{2L-2} \Bigl[  \left|{\ind}\right\rangle \bra{\ind +1}+\ket{\ind +1}\bra{\ind }  +  h_\ind \ket{\ind}\left\langle \ind \right|  \Bigr].
\label{eq:Hmat}
\ee
The two main differences to the ``canonical'' Anderson model lie in the absence of disorder on odd sites and the fact that the $h_\ind$ are identically distributed, but not independent random variables.

\begin{figure}
  \includegraphics[width=.9\columnwidth]{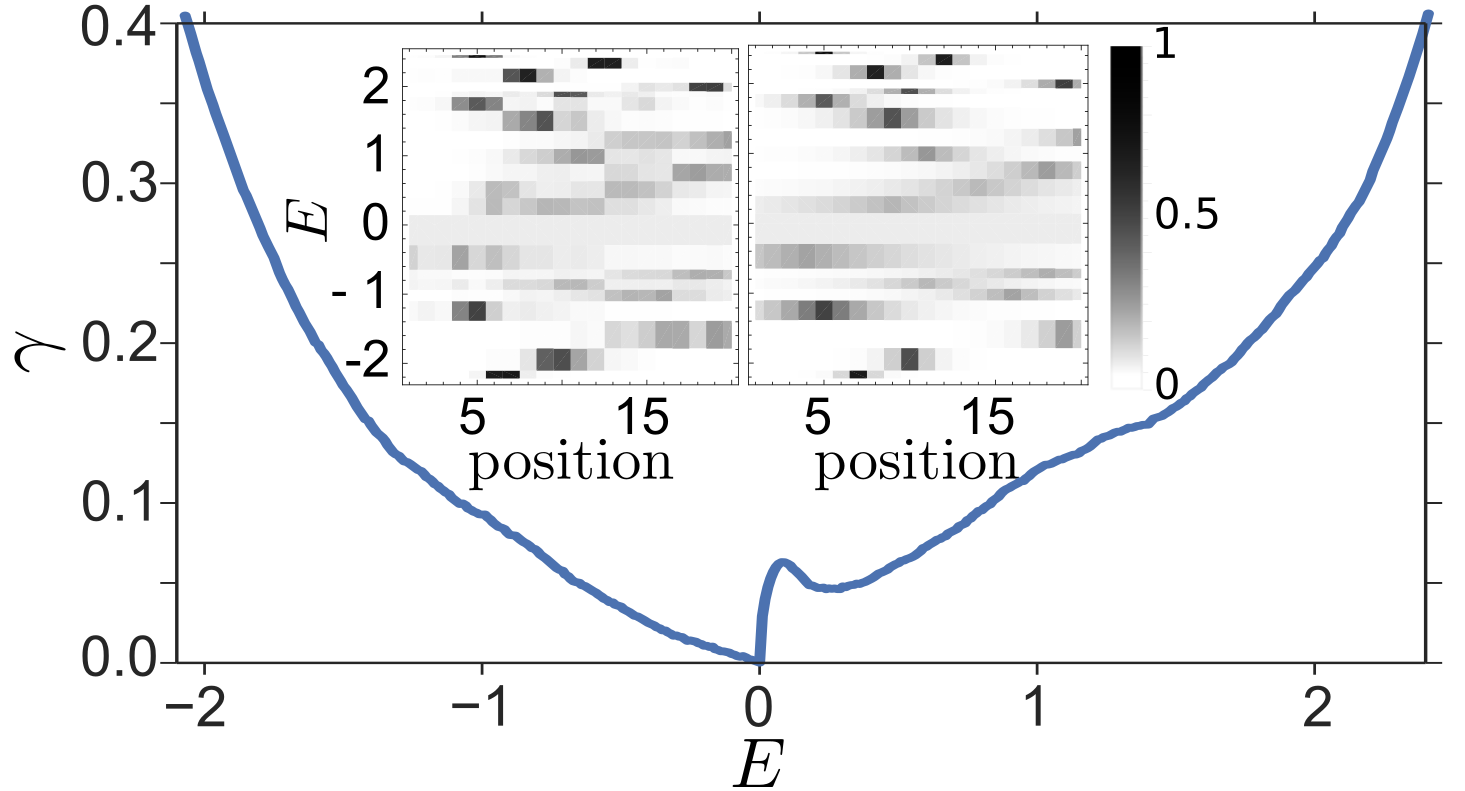}
  \caption{Lyapunov exponent for the one-dimensional Anderson model. All data shown in this figure are obtained with the same parameters employed in the two-atom experiment. In the main figure we report the Lyapunov exponent as a function of the energy $E$ (measured in units of $\Omega/2$). 
  The inset shows a comparison between the shapes of the wave functions obtained from a numerical reconstruction via equation \eqref{eq:rec_step} (left panel) and from the corresponding prediction associated to the Lyapunov exponent (right panel) for a chain of $L = 20$ sites and a specific realization of the disorder. %
  In the right panel the envelopes $\propto \exp\left[-4\gamma(E) \abs{k - k_{\rm max}(E)}\right]$ 
are centered at the position $k_{\rm max}(E)$ at which the corresponding set of excitation probabilities in the left panel reaches its maximum. The factor $4$ in the exponent is half due to considering probabilities instead of amplitudes, half due to the fact that the real chain is approximately half of the length of the one in Fock space.
  }
\label{fig:lyap}
\end{figure}

\begin{figure*}
  \includegraphics[width=1.8\columnwidth]{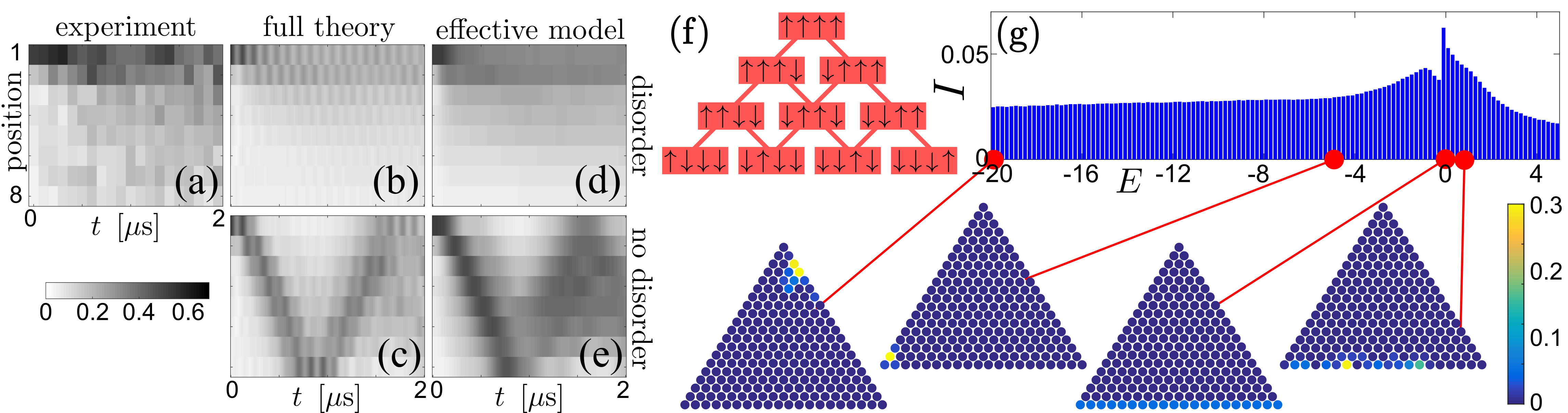}
  \caption{Eight-atom experiment and two-dimensional Anderson model. Panel (a) shows the experimental data for the dynamics of a single excitation averaged over more than 100 realizations. Here, $\ket{\uparrow}=\ket{56D_{3/2},F=3,M=3}$ of $^{87}$Rb, $r_0 = 4.1\;\mu$m, $\Omega = 2\pi \times 2.1$ MHz, $\Delta = -V_{\rm NN} = -2\pi \times 8.4$ MHz and the estimated position uncertainties are $\sigma_1 = 1\;\mu$m and $\sigma_{2,3} = 120$ nm respectively. The data are compared with a numerical integration of the dynamics for both the full Hamiltonian, and the effective 2D Anderson model, with and without disorder (b-e). When disorder is considered the numerical data are averaged over 100 realizations. Panels (b) and (c) are computed using the full Hamiltonian \eqref{eq:H_0}, whereas (d) and (e) are evaluated with an effective 2D Anderson model, as discussed in the main text. In panel (f) the lattice structure of this reduced model is reported for $L=4$ atoms and $N_{\rm cl}=1$. Links are drawn between states connected by one spin flip. 
  Panel (g) shows the inverse participation ratio $I$ as a function of the energy $E$ (measured in units of $\Omega/2$) for a chain of $L=20$ atoms.
  The amplitude of the wave function is reported for four representative states on a lattice whose structure follows the one shown in panel (f). From left to right they display: a state localized in Fock space, but delocalized in real space, a state localized in both, the special state $\ket{\psi_0}$ discussed in the main text and a similar state found for small $E>0$.
  }
\label{fig:4}
\end{figure*}

\section{Localization in the 1D generalized Anderson model}
\label{sec:Anderson 1D}

Henceforth for simplicity we measure all energies and (inverse) times in units of (half) the Rabi frequency, setting $\Omega = 2$.
We approach the problem with a transfer matrix formalism: expressing the quantum state in the restricted Fock basis $\ket{\ind}$, $\ket{\psi} = \sum_\ind a_\ind \ket{\ind}$,
the Schr\"{o}dinger equation $H_{\mathrm{A}} \ket{\psi} =  E \ket{\psi} $ reduces to the recursion equation
\be
	\matb{c} a_{\ind +1} \\ a_\ind \mate =  \matb{cc} E-h_{\ind}  & -1 \\ 1 & 0 \mate \matb{c} a_{\ind} \\ a_{\ind -1} \mate \equiv M_\ind \matb{c} a_{\ind} \\ a_{\ind -1} \mate ,
	\label{eq:rec_step}
\ee
where $M_\ind = M_\ind(E)$ is an energy-dependent transfer matrix which progressively reconstructs the wave function amplitudes from left to right.
The values of $E$ belonging to the spectrum of the Hamiltonian are identified by the boundary conditions $a_{2L}=a_0=0$.

The localization length $l$ can be expressed in terms of the \emph{Lyapunov exponent} \cite{Izrailev1995,Izrailev1999}, 
\be
	\gamma(E) \equiv \lim_{n\to \infty} \frac{1}{n} \log \norm{\prod_{\ind =n}^{1} M_\ind(E)}_{\rm{op}} \equiv l^{-1}
	\label{eq:Lyap}
\ee
(for the existence of the limit see \cite{Furstenberg1960}). The amplitude of a wavefunction corresponding to an eigenvalue $E$ of $H_{\mathrm{A}}$ is concentrated within a region of width $\propto l$. Outside of this region it decays as $\sim \rme{-r/l}$ with the distance $r$. Wavefunctions with $\gamma >0$ are therefore localized, while delocalized states are characterized by $\gamma = 0$.
To illustrate this in our case, we report in Fig.~\ref{fig:lyap} a numerical study of the Lyapunov exponent for a (rather idealized) chain of length $L = 25000$ sites.
We find that $\gamma$ is positive $\forall E \neq 0$, while $\gamma(E = 0) = 0$, signaling the presence of a delocalized state. The asymmetric shape originates from an asymmetry of the distribution of energy displacements between positive and negative values (see Appendix \ref{app:correlations}).

Actually, independently of the realization of the disorder, $E = 0$ is always an eigenvalue of $H_{\mathrm{A}}$ corresponding to the (delocalized) wavefunction $\ket{\psi_0}  = (1/\sqrt{L}) \sum_\ind \sin (\pi \ind /2) \ket{\ind}$, which has nonvanishing components only on states not affected by the disorder. This is in contrast with the standard Anderson model \cite{Anderson1958}, which features full localization, and is instead reminiscent of related works on one dimensional models: the random dimer model \cite{Flores1989,Dunlap1990,Bovier1992,Izrailev1995,DeBievre2000} and the Anderson model in the presence of correlated disorder \cite{Izrailev1999}, both featuring the presence of delocalized states in the spectrum.

The remaining eigenvalues depend instead on the specific realization of the disorder; a numerical analysis for different values of the parameters seems to suggest that all other states are localized ($l < \infty$). In the inset we compare our Lyapunov exponent results with a numerical simulation of a system of size $L=20$. Despite being only well-defined on large scales, the Lyapunov exponent provides in our case reasonable predictions already for relatively small system sizes.

\section{Experiment and localization in the 2D generalized Anderson model}
\label{sec:Anderson 2D}

Turning back to the experiment with Rydberg atoms tightly confined in optical tweezers, we now study experimentally an excitation propagating in a chain of 8 atoms using the setup described in \cite{Labuhn2015}. 
We focus on the evolution of the local densities $\av{n_k(t)}$ starting from a single excitation at one end of the chain $\ket{\psi_{\rm in}} = \ket{\uar \dar \dar\dar\dar\dar\dar\dar}$. 
The results are reported in Fig.~\ref{fig:4}(a) and show no appreciable propagation beyond the second site, indicating suppression of transport.
To make it more evident we compare the experiment with numerical integration of the dynamics for the Hamiltonian (\ref{eq:H_0}). In the presence of disorder, Fig.~\ref{fig:4}(b), the numerical results are comparable with the experimental ones, while the case without randomness, Fig.~\ref{fig:4}(c), clearly features propagation.

In this specific experimental situation (see the caption of Fig.~\ref{fig:4} for details), the condition (ii) of strong next-nearest neighbour blockade, $V_{\rm NNN} \gg \delta V_k$, is not satisfied (note that $V_{\rm NNN} = V_{\rm NN} / 64$). It is thereby possible to grow clusters beyond the two-excitation limit. This breaks the chain-like structure highlighted in Fig.~\ref{fig:2}(b) and gives rise instead, for a single cluster ($N_{\rm cl} = 1$), to a two-dimensional square lattice with $N = L(L+1)/2$ states (sketched in Fig.~\ref{fig:4}(f) for four atoms), as previously found in \cite{Mattioli2015} as well. We remark that the two bottommost rows correspond precisely to the previous one-dimensional chain.
The dynamics on this ``triangle'' of states can then be described by a 2D tight-binding Anderson model similar to Eq.~(\ref{eq:Hmat}) (see Appendix \ref{app:Hilbert} for the derivation). Interestingly, in this regime the chain of atoms can be thought of as a quantum simulator of a synthetic dimension \cite{Boada:2012,Celi:2014,Price:2015,Mancini:2015,Stuhl:2015,Barbarino:2016}; it is also worth mentioning that, increasing the number of clusters ($N_{\rm cl} > 1$), one can go even further and obtain higher-dimensional instances. 
We report in Fig.~\ref{fig:4}(d)-(e) a numerical study of the dynamics for $N_{\rm cl} = 1$ which shows reasonable agreement with both the experiment and the full Hamiltonian dynamics.

These results suggest again the presence of localized states governing the evolution; analogously to the previous case, we focus on the spreading of an eigenstate $\ket{E}$ in the new restricted Fock basis $\ket{\indd}$. 
We quantify this with the inverse participation ratio (IPR) $I  =  (N \sum_\indd \left| \bracket{E}{\indd}\right|^{4})^{-1}$ (first introduced in \cite{Bell1970}). As a measure of localization, the IPR can be easily tested on the two limiting cases: for a state $\ket{E}$ uniformly distributed on the basis ($\abs{\bracket{E}{\indd}} = 1/\sqrt{N}$) one finds the maximal value $I = 1$, whereas for a completely localized state, namely $\ket{E} \equiv \ket{\bar{\indd}}$ corresponding to a single Fock state $\ket{\bar{\indd}}$, one has $I = 1/N$. 
A numerical study of $I$ for $L=20$ atoms and the parameter set employed in the 8-atom experiment is reported in Fig.~\ref{fig:4}(g), where for every realization of the disorder the spectrum is calculated via exact diagonalization. The IPR is then computed for each energy eigenvector and a first average is calculated among levels which end up in the same bin of the histogram. A second average is then applied over all the considered realizations. In general, we observe that the IPR remains rather low on the entire spectrum ($I < 0.1$), signaling that the parameters are in the localized phase.
The form of the IPR indicates the presence of strongly localized states at large energies (both positive and negative), while eigenstates at smaller energies are slightly more spread-out. The central peak links to the presence of the state $\ket{\psi_0} = \ket{E=0}$ encountered above, which is still an exact eigenstate, but only occupies the bottommost row (see example in Fig.~\ref{fig:4}(g)), its IPR being $I = L/N = 2/(L+1)$. This appears to be the most delocalized pattern for the parameter regime considered. The sudden dip on the negative side is due to the absence for $E<0$ of similarly spread-out states on the lowermost rows and will be object of future theoretical investigations. 
It is important to remark that, in contrast to the 1D case, here localization in the Fock space does not necessarily imply localization in real space. In fact, high-energy states might be localized around the tip of the triangle (see example in Fig.~\ref{fig:4}(g)) and encompass Fock states with system-spanning clusters. The present experiment, however, highlights suppressed transfer and thus implies that the initial condition has, for its most part, component on states which are localized in real space as well.

\section{Outlook}
\label{sec:Outlook}
We have shown that the facilitation dynamics in disordered Rydberg lattices is governed by certain classes of tight binding Anderson models. The simplest one is a 1D Anderson model with disorder on every other site for which we have established a thorough connection.
In experimentally relevant parameter regimes we still find inhibition of transport, and interpret it in terms of the physics of a 2D Anderson model with correlated disorder, whose behavior is largely unexplored. 
This connection can be used to shed light on how Fock space localization influences real space localization, which is a subtle and interesting open problem. Our work suggests that this issue can be now addressed experimentally with Rydberg atoms and provides theoretical grounds for future investigations.

\section{Acknowledgments}
IL thanks Juan P. Garrahan for fruitful discussions. The research leading to these results has received funding from the European Research Council under the European Union's Seventh Framework Programme (FP/2007-2013) / ERC Grant Agreement No. 335266 (ESCQUMA), the EU-FET Grant No. 512862 (HAIRS), the H2020-FETPROACT-2014 Grant No.640378 (RYSQ), and EPSRC Grant No. EP/M014266/1 and by the R\'egion Ile-de-France in the framework of DIM Nano-K.


\input{andersonMBL_arXiv_v1_bib.bbl}

\appendix

\section{Approximate Gaussian distribution of the atomic positions}
\label{app:distribution}

Here we recall how the Gaussian distribution of the atomic positions arises. As a first approximation, we assume the motional degrees of freedom to be classical, so that we can describe the position of the atom by the Boltzmann distribution $f(\mathbf{r}, \mathbf{p}) = \exp \lt -\beta H_{\rm motion}(\mathbf{r}, \mathbf{p})  \rt$. For low enough temperatures, the atoms have only access to the harmonic part of the potential and $H_{\rm motion} (\mathbf{r}, \mathbf{p}) \approx \sum_i p_i^2 / (2m) + (m/2) \sum_i \omega_i^2 r_i^2$. The distribution of the positions $p_{\rm pos} = (\int \rmd^3 p \, f) / (\int \rmd^3 p \, \rmd^3 r \, f)$ can be read off directly and is a Gaussian with zero mean and variances $\sigma_i^2 = 1/(m\omega_i^2 \beta)$.
The complete three-dimensional distribution is then simply a product of $p_{\rm pos}(x_i)$ along the three directions. 
For an atom in a trap centered at position $k\mathbf{r}_0 = (0,0,kr_0)$ with $k$ an integer, it is straightforwardly generalized to
\be
	p^{(k)}_{\rm pos}(\mathbf{r}) = \frac{1}{\lt 2\pi \rt^{3/2} \sigma_1 \sigma_2 \sigma_3} 
	\rme{-  \frac{r_1^2}{2\sigma_1^2} -\frac{r_2^2}{2\sigma_2^2} -\frac{(r_3 - k r_0)^2}{2\sigma_3^2} }.
	\label{eq:distr0}
\ee
We remark that the indices in the expression above distinguish between Cartesian components only, e.g $r_1$ and $r_2$ are the components along $x$ and $y$ of the same atomic position. In the following, whenever necessary to display both, the trap index will always appear before the component one, e.g., $r_{k,i}$ is the $i$-th component of the $k$-th atom's position.

\section{Correlation of the distances and typical interaction displacements}
\label{app:correlations}

In this section we explain how the independent atomic positions lead to correlated inter-atomic distances and, in turn, to correlated energy fluctuations. We comment on the respective probability distributions.

In our numerical simulations, each atomic position $\mathbf{r}_k$ is independently generated according to the distribution \eqref{eq:distr0} relative to its own trap. As explained in the main text, the nearest-neighbour differences $\mathbf{d}_k = \mathbf{r}_{k+1} - \mathbf{r}_k = (d_{k,1}, d_{k,2}, d_{k,3})$ are not independent - for example, both $\mathbf{d}_1$ and $\mathbf{d}_2$ depend on the position of the second atom. The joint distribution of $\mathbf{d}_k$s can be obtained from the atomic positions distribution as
\begin{widetext}
\be
\begin{split}
	p_{\rm diff}(\mathbf{d}_1 , \ldots , \mathbf{d}_{L-1}) = \int \lqq \prodl{k=1}{L} \rmd^3 r_k \, p^{(k)}_{\rm pos}(\mathbf{r}_k) \rqq  
	\lqq  \prodl{k'=1}{L-1} \delta^{(3)} \lt  \mathbf{d}_{k'} - \lt \mathbf{r}_{k'+1} - \mathbf{r}_{k'} \rt  \rt \rqq = \\
	= \lqq  \frac{1}{\sqrt{L} \lt \sqrt{2\pi } \rt^{L-1}}  \rqq^3  \lt \sigma_1 \sigma_2 \sigma_3 \rt^{1-L}  \rme{- \ha \sum_{k,q} \lqq \frac{1}{\sigma_1^2}  d_{k,1} A_{kq} d_{q,1}  - \frac{1}{\sigma_2^2}  d_{k,2} A_{kq} d_{q,2}  - \frac{1}{\sigma_3^2}  (d_{k,3} - r_0) A_{kq} (d_{q,3} - r_0)  \rqq } ,   
	\label{eq:pdiff}
\end{split}
\ee
\end{widetext}
where $A_{kq} = L - \max(k,q) - (L-k)(L-q)/L = (L - \max(k,q)) \min(k,q) / L$ is a symmetric real matrix. From here, one can determine the correlation properties of the distances: the correlation matrix $C = A^{-1}$ is a tridiagonal matrix \cite{MatInverse}
\be
C =  \matb{ccccc}    
	2 & -1 & 0 & 0 &   \\
	-1 & 2 & -1 & 0 &   \\
	0 & -1 & 2 & -1 &  \cdots  \\
	0 & 0 & -1 & 2 &    \\
	  &   &  \vdots &    & \ddots
\mate
\ee
implying e.g. $\av{d_{k,3} d_{q,3}}_c \equiv \av{d_{k,3} d_{q,3}} - \av{d_{k,3}} \av{ d_{q,3}} = \sigma_3^2 \lt  2 \delta_{k,q} - \delta_{k,q+1} - \delta_{k,q-1}  \rt$. It confirms the expected result, namely that contiguous distances are \mbox{(anti-)correlated}. This comes from the simple fact that, considering three atoms, moving the middle atom closer to the first one brings it further away from the last one.

\begin{figure}
  \includegraphics[width=0.7\columnwidth]{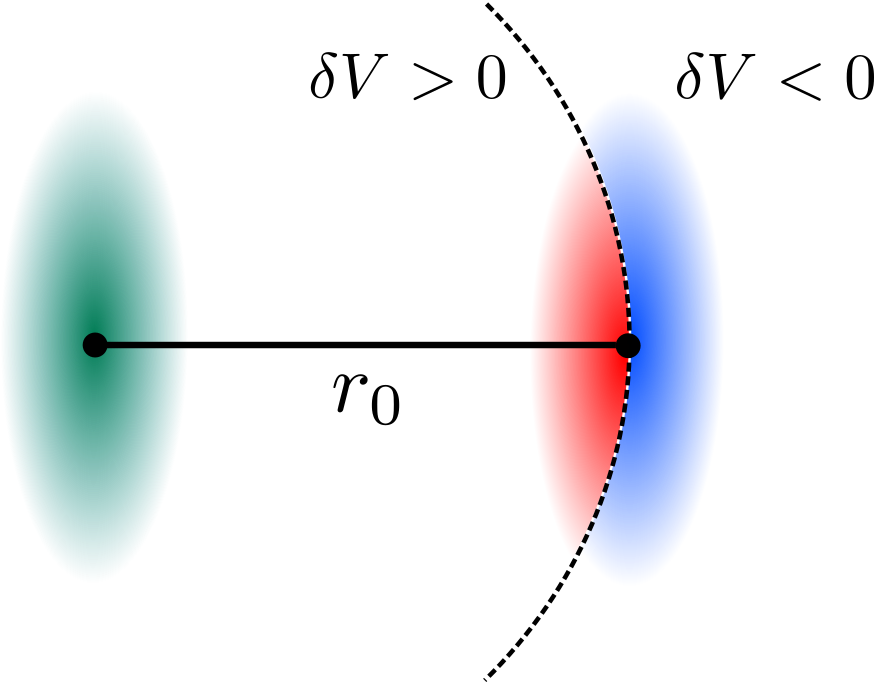}
  \caption{Asymmetry of the energy displacements. Here an excitation is present in the center of the leftmost trap (in green). The dashed line indicates the facilitation shell, i.e., the sphere of points where $V(r_0) = - \Delta$. For repulsive interactions, the red portion of the second trap corresponds to the domain where the energy displacement is positive ($\delta V > 0$), whereas the opposite ($\delta V < 0$) holds for the blue one. It is then apparent that the volume covered by the blue portion is larger than the volume of the red one, yielding the mentioned bias towards negative values.  
  }
\label{fig:M1}
\end{figure}

As mentioned in the main text, the asymmetric profiles of both the Lyapunov exponent (for the 1D case) and the inverse participation ratio (for the 2D case), stem from the asymmetry of the distribution $p_{\rm int} (\delta V)$ of energy displacements. For anisotropic traps ($\sigma_i \neq \sigma_j$) there is no closed formula for $p_{\rm int}$. However, considering for instance repulsive interactions ($V(r) > 0$), the bias towards negative values ($\delta V <0$) can still be understood simply by analyzing the geometry of the setup: in Fig.~\ref{fig:M1} we display two neighboring traps. The facilitation radius $r_0$ corresponds to the distance at which the detuning $\Delta$ exactly cancels the interaction $V(r_0)$ and thus separates the regime $\delta V > 0$ (inside, $d < r_0$, red area in the figure) from $\delta V > 0$ (outside, $d > r_0$, blue area in the figure). It then becomes apparent that the former includes a smaller portion of the second trap than the latter. In other words, setting as a first approximation the first atom in the center of its trap, the placement of the second one will more likely yield a distance $d > r_0$ than the converse. For attractive interactions, the signs change and the bias will be towards positive values. 

The typical energy displacement can also be estimated by simple considerations: taking two neighboring atoms at average separation $\mathbf{r}_0 = (0,0,r_0)$ and standard deviation (of the distance between them) $\overline{\delta r} = \sqrt{\av{d^2} - \av{d}^2} \approx \sqrt{2}\sigma_3$, we define
\be
	\overline{\delta V} = \left| \frac{\partial V}{\partial r} \right| \overline{\delta r} = 6 \left| V(r_0)\right| \frac{\overline{\delta r}}{r_0}.
\ee
We emphasize that we only include here the contribution $\sigma_3$, which is the only one acting to first order in $\sigma_{1,2,3}/d $. This yields a reasonable lower bound on $\overline{\delta V}$.

For the set of parameters used in the two-atoms experiment ($\sigma_{3} = 0.12 \,\mu{\rm m}$, $r_0=14.2\;\mu \rm{m}$, $V(r_0) = 2 \pi \times 8.9 \, {\rm MHz}$) we find $\overline{\delta V} \approx 2 \pi \times 0.64 \,{\rm MHz}$. 
For the eight-atoms experiment ($\sigma_{3} = 0.12 \,\mu{\rm m}$, $r_0=4.1\;\mu \rm{m}$, $V(r_0) = 2 \pi \times 8.4 \, {\rm  MHz}$) we obtain  $\overline{\delta V} \approx 2 \pi \times 2.1 \,{\rm MHz}$. This value is to be compared with the Rabi frequency $\Omega \approx 2\pi \times 2.1 \,{\rm MHz}$ and confirms the relevance of the disorder for the propagation of excitations in this setup.

\section{Hilbert space reductions and restricted Hamiltonians}
\label{app:Hilbert}

Here we provide the detailed derivation of the effective 1D and 2D Hamiltonians. For the reader's convenience, we recall here from the main text the original Hamiltonian 
\be
	H =  \sum_k   \lqq  \frac{\Omega}{2} \sigma^x_k +   \Delta n_k   + \sum_{l>k} V(\abs{\mathbf{r}_k - \mathbf{r}_l} )   n_k n_l\rqq
	\label{eq:H_0 methods}
\ee
of the model. For simplicity, we are going to neglect all interactions beyond next-nearest neighbors (NNN) (for the parameters above, e.g., $\abs{V(3r_0) / \Omega} \sim 10^{-3}$), so that the second sum above can be restricted to $l = k+1 , k+2$. Second, the relative displacement between NNNs is suppressed by a factor $2^6 = 64$ with respect to the noise between nearest neighbors and can therefore also be discarded. After these basic approximations, $H$ takes the form
\be
\begin{split}
	H = &  \sum_k   \Bigl[  \frac{\Omega}{2} \sigma^x_k +   \Delta n_k   +  (V_{\rm NN} + \delta V_k)   n_k n_{k+1} + \Bigr. \\
	+ & \Bigl. V_{\rm NNN}  n_k n_{k+2} \Bigr] = \\
	= & \sum_k   \Bigl[  \frac{\Omega}{2} \sigma^x_k +   \Delta n_k (1 - n_{k+1})  + \delta V_k   n_k n_{k+1} + \Bigr. \\
	+ & \Bigl.  V_{\rm NNN}  n_k n_{k+2} \Bigr]
	\label{eq:H_0 2 methods}
\end{split}
\ee
where we used the facilitation constraint $V_{\rm NN} = - \Delta$. Note that the sum runs over $k = 1 \ldots L$ and, for later convenience, we fix four auxiliary variables $n_{-1} = n_{0} = n_{L+1} = n_{L+2} \equiv 0$. We now enforce condition (i) $\Delta \gg \Omega$. This implies that spin flips are strongly suppressed if not in the presence of a single excited neighbor; we further approximate our Hamiltonian by making this a hard constraint. In other words, the transitions $\ket{\dar \dar \dar} \leftrightarrow \ket{\dar \uar \dar}$ and $\ket{\uar \dar \uar} \leftrightarrow \ket{\uar \uar \uar}$ are prohibited. If we now define a ``cluster'' as an uninterrupted sequence of $\uar$ spins (for instance, the state $\ket{\dar \boxed{\uar \uar} \dar \boxed{\uar} \dar \boxed{\uar \uar \uar}}$ has three highlighted clusters), we see that these structures cannot appear or disappear, nor can they merge or split. Hence, as pointed out in \cite{Mattioli2015} as well, the number $N_{\rm cl}$ of these clusters is conserved. In particular, having fixed $n_{L+1} = 0$, the number of clusters corresponds to the number of right kinks $\ket{\uar \dar}$, i.e., $N_{\rm cl} = \sum_{k=1}^L n_k (1-n_{k+1})$. The Hamiltonian now reads
\be
H =   \Delta N_{\rm cl} + \sum_k   \Bigl[  \frac{\Omega}{2} \sigma^x_k P^{(i)}_k + \delta V_k   n_k n_{k+1} + 	 V_{\rm NNN}  n_k n_{k+2} \Bigr]
\ee
with the projector $P^{(i)}_k = n_{k-1} + n_{k+1} - 2 n_{k-1} n_{k+1}$. If we consider now the special case $N_{\rm cl} = 1$, we notice that the states with a single cluster can be labeled by two indices: the starting position of the cluster ($\mu = 1 \ldots L$) and the ending one $\nu = \mu \ldots L$. In order to enforce the condition $\nu \geq \mu$ and avoid spurious boundary terms, we formally use the projector $\Theta \ket{\mu, \nu} = \theta(\nu - \mu) \theta (\mu) \theta(\nu) \theta (L - \mu) \theta(L - \nu) \ket{\mu, \nu}$ on the valid states, where $\theta (x )$ is the Heaviside step function ($\theta(x \geq 0) = 1$ and $\theta (x< 0) = 0$). Since clusters only grow/shrink at the edges, the Hamiltonian can be recast in the form
\begin{subequations}
\begin{align}
	H_B & = \Omega \, \Theta H_B'  \Theta  \quad \text{with} \\
	H_B' & = \sum_{\mu, \nu = 1}^L \Bigl[  \ha \lt \ket{\mu, \nu} \bra{\mu + 1 , \nu}  + \ket{\mu, \nu} \bra{\mu  , \nu + 1}  + h.c. \rt   + \Bigr. \nol 
	&  \Bigl.   \lt \delta v_{\mu \nu}  + (\nu - \mu  - 2) \theta (\nu - \mu  - 2) \frac{V_{\rm NNN}}{\Omega}   \rt  \proj{\mu , \nu}  \Bigr]  ,
\end{align}
\end{subequations}
where $\delta v_{\mu \nu} = \sum_{k = \mu}^{\nu - 1} \delta V_k / \Omega$ and for simplicity we subtracted the additive constant $\Delta$. In this notation, one can regard $H_B$ as a hopping Hamiltonian on half a square lattice (since we take $\nu \geq \mu$), as reported in the main text. Each site feels a random potential $\delta v_{\mu \nu}$ and a deterministic one originating from the NNN interactions (provided of course, that there are more than two $\uar$ spins in the cluster). It is therefore reminiscent of a 2D Anderson problem, the main difference being in the peculiar form of the noise, which appears as the sum of at most $L-1$ random variables and makes it non-trivially correlated between different sites.

The 1D Anderson-like model we introduce in our main text is obtained when condition (ii) $V_{\rm NNN} \gg \Omega$ also holds. By approximating this as a hard constraint (i.e., assuming the limit $V_{\rm NNN} / \Omega \to \infty$) the number of next-nearest-neighboring excitations $N_{\rm NNN}$ becomes a conserved quantity. The Hamiltonian then reads
\be
H =   \Delta N_{\rm cl} + N_{\rm NNN} V_{\rm NNN} + \sum_k   \Bigl[  \frac{\Omega}{2} \sigma^x_k P^{(i)}_k P^{(ii)}_k + \delta V_k   n_k n_{k+1}  \Bigr],
\label{eq:H2}
\ee
with the additional projector $P^{(ii)}_k = (1- n_{k-2}) (1- n_{k+2})$. Note that under these conditions spins neighboring a pair of excitations cannot flip (e.g., $\ket{\uar \uar \dar} \leftrightarrow \ket{\uar \uar \uar}$ is suppressed). Similarly, different clusters cannot grow to a distance smaller than two now (i.e., transitions such as $\ket{\uar \dar \dar \uar} \leftrightarrow \ket{\uar \uar \dar \uar}$ are prohibited as well). This means that any longer-than-two cluster is a stable local configuration (i.e., invariant under the dynamics generated by \eqref{eq:H2}) which cuts the chain of atoms in two dynamically-disconnected parts. Each of these parts can be read as a subsystem subject to the same Hamiltonian \eqref{eq:H2} but with lower $N_{\rm NNN}$. Therefore, the analysis can be restricted, without conceptual loss, to the case $N_{\rm NNN} = 0$. The description becomes particularly simple for $N_{\rm cl} = 1$, since the states can be labeled simply by $b = 2p-1$, with $p$ the position of the ``center of mass'' of the clusters:
\begin{align}
	&\ket{\uar \dar \dar \dar \ldots } \equiv \ket{1} \\
	&\ket{\uar \uar \dar \dar \ldots} \equiv \ket{2 \frac{3}{2} - 1} = \ket{2} \\
	&\ket{\dar \uar \dar \dar \ldots} \equiv \ket{3} \\
	&\ldots
\end{align} 
The advantage of this labeling is that the states are now sequentially connected by the Hamiltonian, i.e., $\bra{b} H \ket{b'} \neq 0 \Leftrightarrow (b - b') = 0, \pm 1$ and thus naturally define a chain. Subtracting the additive constant $\Delta$, one then finds again equation \eqref{eq:Hmat} of the main text, i.e.,
\begin{equation}
H_{\mathrm{A}}=\frac{\Omega}{2}\sum_{\ind =1}^{2L-2} \Bigl[ \left|{\ind}\right\rangle \bra{\ind +1}+\ket{\ind +1}\bra{\ind }  +  h_\ind \ket{\ind}\left\langle \ind \right|  \Bigr] ,
\label{eq:Hmat methods}
\end{equation}
where
\be
	h_\ind  = \sysb{lll} 0 & & \text{(if $b$ odd)} \\[2mm] 2 \delta V_{b/2} / \Omega & & \text{(if $b$ even)}.  \syse
\ee

\clearpage
\end{document}

%% file: andersonMBL_arXiv_v1_bib.bbl
%

%% file: andersonMBL_arXiv_v1.bbl
\begin{thebibliography}{47}%
\makeatletter
\providecommand \@ifxundefined [1]{%
 \@ifx{#1\undefined}
}%
\providecommand \@ifnum [1]{%
 \ifnum #1\expandafter \@firstoftwo
 \else \expandafter \@secondoftwo
 \fi
}%
\providecommand \@ifx [1]{%
 \ifx #1\expandafter \@firstoftwo
 \else \expandafter \@secondoftwo
 \fi
}%
\providecommand \natexlab [1]{#1}%
\providecommand \enquote  [1]{``#1''}%
\providecommand \bibnamefont  [1]{#1}%
\providecommand \bibfnamefont [1]{#1}%
\providecommand \citenamefont [1]{#1}%
\providecommand \href@noop [0]{\@secondoftwo}%
\providecommand \href [0]{\begingroup \@sanitize@url \@href}%
\providecommand \@href[1]{\@@startlink{#1}\@@href}%
\providecommand \@@href[1]{\endgroup#1\@@endlink}%
\providecommand \@sanitize@url [0]{\catcode `\\12\catcode `\$12\catcode
  `\&12\catcode `\#12\catcode `\^12\catcode `\_12\catcode `\%12\relax}%
\providecommand \@@startlink[1]{}%
\providecommand \@@endlink[0]{}%
\providecommand \url  [0]{\begingroup\@sanitize@url \@url }%
\providecommand \@url [1]{\endgroup\@href {#1}{\urlprefix }}%
\providecommand \urlprefix  [0]{URL }%
\providecommand \Eprint [0]{\href }%
\providecommand \doibase [0]{http://dx.doi.org/}%
\providecommand \selectlanguage [0]{\@gobble}%
\providecommand \bibinfo  [0]{\@secondoftwo}%
\providecommand \bibfield  [0]{\@secondoftwo}%
\providecommand \translation [1]{[#1]}%
\providecommand \BibitemOpen [0]{}%
\providecommand \bibitemStop [0]{}%
\providecommand \bibitemNoStop [0]{.\EOS\space}%
\providecommand \EOS [0]{\spacefactor3000\relax}%
\providecommand \BibitemShut  [1]{\csname bibitem#1\endcsname}%
\let\auto@bib@innerbib\@empty
\bibitem [{\citenamefont {Anderson}(1958)}]{Anderson1958}%
  \BibitemOpen
  \bibfield  {author} {\bibinfo {author} {\bibfnamefont {P.~W.}\ \bibnamefont
  {Anderson}},\ }\href {\doibase 10.1103/PhysRev.109.1492} {\bibfield
  {journal} {\bibinfo  {journal} {Phys. Rev.}\ }\textbf {\bibinfo {volume}
  {109}},\ \bibinfo {pages} {1492} (\bibinfo {year} {1958})}\BibitemShut
  {NoStop}%
\bibitem [{\citenamefont {Mott}\ and\ \citenamefont {Twose}(1961)}]{Mott1961}%
  \BibitemOpen
  \bibfield  {author} {\bibinfo {author} {\bibfnamefont {N.}~\bibnamefont
  {Mott}}\ and\ \bibinfo {author} {\bibfnamefont {W.}~\bibnamefont {Twose}},\
  }\href {\doibase 10.1080/00018736100101271} {\bibfield  {journal} {\bibinfo
  {journal} {Advances in Physics}\ }\textbf {\bibinfo {volume} {10}},\ \bibinfo
  {pages} {107} (\bibinfo {year} {1961})}\BibitemShut {NoStop}%
\bibitem [{\citenamefont {Ishii}(1973)}]{Ishii1973}%
  \BibitemOpen
  \bibfield  {author} {\bibinfo {author} {\bibfnamefont {K.}~\bibnamefont
  {Ishii}},\ }\href@noop {} {\bibfield  {journal} {\bibinfo  {journal}
  {Progress of Theoretical Physics Supplement}\ }\textbf {\bibinfo {volume}
  {53}},\ \bibinfo {pages} {77} (\bibinfo {year} {1973})}\BibitemShut {NoStop}%
\bibitem [{\citenamefont {Cutler}\ and\ \citenamefont
  {Mott}(1969)}]{Cutler:1969}%
  \BibitemOpen
  \bibfield  {author} {\bibinfo {author} {\bibfnamefont {M.}~\bibnamefont
  {Cutler}}\ and\ \bibinfo {author} {\bibfnamefont {N.~F.}\ \bibnamefont
  {Mott}},\ }\href@noop {} {\bibfield  {journal} {\bibinfo  {journal} {Physical
  Review}\ }\textbf {\bibinfo {volume} {181}},\ \bibinfo {pages} {1336}
  (\bibinfo {year} {1969})}\BibitemShut {NoStop}%
\bibitem [{\citenamefont {Billy}\ \emph {et~al.}(2008)\citenamefont {Billy},
  \citenamefont {Josse}, \citenamefont {Zuo}, \citenamefont {Bernard},
  \citenamefont {Hambrecht}, \citenamefont {Lugan}, \citenamefont
  {Cl{\'e}ment}, \citenamefont {Sanchez-Palencia}, \citenamefont {Bouyer},\
  and\ \citenamefont {Aspect}}]{Billy:2008}%
  \BibitemOpen
  \bibfield  {author} {\bibinfo {author} {\bibfnamefont {J.}~\bibnamefont
  {Billy}}, \bibinfo {author} {\bibfnamefont {V.}~\bibnamefont {Josse}},
  \bibinfo {author} {\bibfnamefont {Z.}~\bibnamefont {Zuo}}, \bibinfo {author}
  {\bibfnamefont {A.}~\bibnamefont {Bernard}}, \bibinfo {author} {\bibfnamefont
  {B.}~\bibnamefont {Hambrecht}}, \bibinfo {author} {\bibfnamefont
  {P.}~\bibnamefont {Lugan}}, \bibinfo {author} {\bibfnamefont
  {D.}~\bibnamefont {Cl{\'e}ment}}, \bibinfo {author} {\bibfnamefont
  {L.}~\bibnamefont {Sanchez-Palencia}}, \bibinfo {author} {\bibfnamefont
  {P.}~\bibnamefont {Bouyer}}, \ and\ \bibinfo {author} {\bibfnamefont
  {A.}~\bibnamefont {Aspect}},\ }\href@noop {} {\bibfield  {journal} {\bibinfo
  {journal} {Nature}\ }\textbf {\bibinfo {volume} {453}},\ \bibinfo {pages}
  {891} (\bibinfo {year} {2008})}\BibitemShut {NoStop}%
\bibitem [{\citenamefont {Roati}\ \emph {et~al.}(2008)\citenamefont {Roati},
  \citenamefont {D’Errico}, \citenamefont {Fallani}, \citenamefont {Fattori},
  \citenamefont {Fort}, \citenamefont {Zaccanti}, \citenamefont {Modugno},
  \citenamefont {Modugno},\ and\ \citenamefont {Inguscio}}]{Roati:2008}%
  \BibitemOpen
  \bibfield  {author} {\bibinfo {author} {\bibfnamefont {G.}~\bibnamefont
  {Roati}}, \bibinfo {author} {\bibfnamefont {C.}~\bibnamefont {D’Errico}},
  \bibinfo {author} {\bibfnamefont {L.}~\bibnamefont {Fallani}}, \bibinfo
  {author} {\bibfnamefont {M.}~\bibnamefont {Fattori}}, \bibinfo {author}
  {\bibfnamefont {C.}~\bibnamefont {Fort}}, \bibinfo {author} {\bibfnamefont
  {M.}~\bibnamefont {Zaccanti}}, \bibinfo {author} {\bibfnamefont
  {G.}~\bibnamefont {Modugno}}, \bibinfo {author} {\bibfnamefont
  {M.}~\bibnamefont {Modugno}}, \ and\ \bibinfo {author} {\bibfnamefont
  {M.}~\bibnamefont {Inguscio}},\ }\href@noop {} {\bibfield  {journal}
  {\bibinfo  {journal} {Nature}\ }\textbf {\bibinfo {volume} {453}},\ \bibinfo
  {pages} {895} (\bibinfo {year} {2008})}\BibitemShut {NoStop}%
\bibitem [{\citenamefont {Semeghini}\ \emph {et~al.}(2015)\citenamefont
  {Semeghini}, \citenamefont {Landini}, \citenamefont {Castilho}, \citenamefont
  {Roy}, \citenamefont {Spagnolli}, \citenamefont {Trenkwalder}, \citenamefont
  {Fattori}, \citenamefont {Inguscio},\ and\ \citenamefont
  {Modugno}}]{Semeghini:2015}%
  \BibitemOpen
  \bibfield  {author} {\bibinfo {author} {\bibfnamefont {G.}~\bibnamefont
  {Semeghini}}, \bibinfo {author} {\bibfnamefont {M.}~\bibnamefont {Landini}},
  \bibinfo {author} {\bibfnamefont {P.}~\bibnamefont {Castilho}}, \bibinfo
  {author} {\bibfnamefont {S.}~\bibnamefont {Roy}}, \bibinfo {author}
  {\bibfnamefont {G.}~\bibnamefont {Spagnolli}}, \bibinfo {author}
  {\bibfnamefont {A.}~\bibnamefont {Trenkwalder}}, \bibinfo {author}
  {\bibfnamefont {M.}~\bibnamefont {Fattori}}, \bibinfo {author} {\bibfnamefont
  {M.}~\bibnamefont {Inguscio}}, \ and\ \bibinfo {author} {\bibfnamefont
  {G.}~\bibnamefont {Modugno}},\ }\href@noop {} {\bibfield  {journal} {\bibinfo
   {journal} {Nature Physics}\ }\textbf {\bibinfo {volume} {11}},\ \bibinfo
  {pages} {554} (\bibinfo {year} {2015})}\BibitemShut {NoStop}%
\bibitem [{\citenamefont {Liao}\ \emph {et~al.}(2015)\citenamefont {Liao},
  \citenamefont {Ou}, \citenamefont {Feng}, \citenamefont {Yang}, \citenamefont
  {Lin}, \citenamefont {Yang}, \citenamefont {Wu}, \citenamefont {He},
  \citenamefont {Ma}, \citenamefont {Xue},\ and\ \citenamefont
  {Li}}]{Liao:2015}%
  \BibitemOpen
  \bibfield  {author} {\bibinfo {author} {\bibfnamefont {J.}~\bibnamefont
  {Liao}}, \bibinfo {author} {\bibfnamefont {Y.}~\bibnamefont {Ou}}, \bibinfo
  {author} {\bibfnamefont {X.}~\bibnamefont {Feng}}, \bibinfo {author}
  {\bibfnamefont {S.}~\bibnamefont {Yang}}, \bibinfo {author} {\bibfnamefont
  {C.}~\bibnamefont {Lin}}, \bibinfo {author} {\bibfnamefont {W.}~\bibnamefont
  {Yang}}, \bibinfo {author} {\bibfnamefont {K.}~\bibnamefont {Wu}}, \bibinfo
  {author} {\bibfnamefont {K.}~\bibnamefont {He}}, \bibinfo {author}
  {\bibfnamefont {X.}~\bibnamefont {Ma}}, \bibinfo {author} {\bibfnamefont
  {Q.-K.}\ \bibnamefont {Xue}}, \ and\ \bibinfo {author} {\bibfnamefont
  {Y.}~\bibnamefont {Li}},\ }\href {\doibase 10.1103/PhysRevLett.114.216601}
  {\bibfield  {journal} {\bibinfo  {journal} {Phys. Rev. Lett.}\ }\textbf
  {\bibinfo {volume} {114}},\ \bibinfo {pages} {216601} (\bibinfo {year}
  {2015})}\BibitemShut {NoStop}%
\bibitem [{\citenamefont {Bitter}\ and\ \citenamefont
  {Milner}(2016)}]{Bitter:2016}%
  \BibitemOpen
  \bibfield  {author} {\bibinfo {author} {\bibfnamefont {M.}~\bibnamefont
  {Bitter}}\ and\ \bibinfo {author} {\bibfnamefont {V.}~\bibnamefont
  {Milner}},\ }\href@noop {} {\bibfield  {journal} {\bibinfo  {journal} {arXiv
  preprint arXiv:1603.06918}\ } (\bibinfo {year} {2016})}\BibitemShut {NoStop}%
\bibitem [{\citenamefont {Altshuler}\ \emph {et~al.}(1997)\citenamefont
  {Altshuler}, \citenamefont {Gefen}, \citenamefont {Kamenev},\ and\
  \citenamefont {Levitov}}]{Altshuler1997}%
  \BibitemOpen
  \bibfield  {author} {\bibinfo {author} {\bibfnamefont {B.~L.}\ \bibnamefont
  {Altshuler}}, \bibinfo {author} {\bibfnamefont {Y.}~\bibnamefont {Gefen}},
  \bibinfo {author} {\bibfnamefont {A.}~\bibnamefont {Kamenev}}, \ and\
  \bibinfo {author} {\bibfnamefont {L.~S.}\ \bibnamefont {Levitov}},\ }\href
  {\doibase 10.1103/PhysRevLett.78.2803} {\bibfield  {journal} {\bibinfo
  {journal} {Phys. Rev. Lett.}\ }\textbf {\bibinfo {volume} {78}},\ \bibinfo
  {pages} {2803} (\bibinfo {year} {1997})}\BibitemShut {NoStop}%
\bibitem [{\citenamefont {Gornyi}\ \emph {et~al.}(2005)\citenamefont {Gornyi},
  \citenamefont {Mirlin},\ and\ \citenamefont {Polyakov}}]{Gornyi2005}%
  \BibitemOpen
  \bibfield  {author} {\bibinfo {author} {\bibfnamefont {I.~V.}\ \bibnamefont
  {Gornyi}}, \bibinfo {author} {\bibfnamefont {A.~D.}\ \bibnamefont {Mirlin}},
  \ and\ \bibinfo {author} {\bibfnamefont {D.~G.}\ \bibnamefont {Polyakov}},\
  }\href {\doibase 10.1103/PhysRevLett.95.206603} {\bibfield  {journal}
  {\bibinfo  {journal} {Phys. Rev. Lett.}\ }\textbf {\bibinfo {volume} {95}},\
  \bibinfo {pages} {206603} (\bibinfo {year} {2005})}\BibitemShut {NoStop}%
\bibitem [{\citenamefont {Basko}\ \emph {et~al.}(2006)\citenamefont {Basko},
  \citenamefont {Aleiner},\ and\ \citenamefont {Altshuler}}]{Basko2006}%
  \BibitemOpen
  \bibfield  {author} {\bibinfo {author} {\bibfnamefont {D.}~\bibnamefont
  {Basko}}, \bibinfo {author} {\bibfnamefont {I.}~\bibnamefont {Aleiner}}, \
  and\ \bibinfo {author} {\bibfnamefont {B.}~\bibnamefont {Altshuler}},\ }\href
  {\doibase http://dx.doi.org/10.1016/j.aop.2005.11.014} {\bibfield  {journal}
  {\bibinfo  {journal} {Annals of Physics}\ }\textbf {\bibinfo {volume}
  {321}},\ \bibinfo {pages} {1126 } (\bibinfo {year} {2006})}\BibitemShut
  {NoStop}%
\bibitem [{\citenamefont {Oganesyan}\ and\ \citenamefont
  {Huse}(2007)}]{Oganesyan2007}%
  \BibitemOpen
  \bibfield  {author} {\bibinfo {author} {\bibfnamefont {V.}~\bibnamefont
  {Oganesyan}}\ and\ \bibinfo {author} {\bibfnamefont {D.~A.}\ \bibnamefont
  {Huse}},\ }\href {\doibase 10.1103/PhysRevB.75.155111} {\bibfield  {journal}
  {\bibinfo  {journal} {Phys. Rev. B}\ }\textbf {\bibinfo {volume} {75}},\
  \bibinfo {pages} {155111} (\bibinfo {year} {2007})}\BibitemShut {NoStop}%
\bibitem [{\citenamefont {Pal}\ and\ \citenamefont {Huse}(2010)}]{Pal2010}%
  \BibitemOpen
  \bibfield  {author} {\bibinfo {author} {\bibfnamefont {A.}~\bibnamefont
  {Pal}}\ and\ \bibinfo {author} {\bibfnamefont {D.~A.}\ \bibnamefont {Huse}},\
  }\href {\doibase 10.1103/PhysRevB.82.174411} {\bibfield  {journal} {\bibinfo
  {journal} {Phys. Rev. B}\ }\textbf {\bibinfo {volume} {82}},\ \bibinfo
  {pages} {174411} (\bibinfo {year} {2010})}\BibitemShut {NoStop}%
\bibitem [{\citenamefont {Yao}\ \emph {et~al.}(2014)\citenamefont {Yao},
  \citenamefont {Laumann}, \citenamefont {Gopalakrishnan}, \citenamefont
  {Knap}, \citenamefont {M{\"u}ller}, \citenamefont {Demler},\ and\
  \citenamefont {Lukin}}]{Yao2014}%
  \BibitemOpen
  \bibfield  {author} {\bibinfo {author} {\bibfnamefont {N.~Y.}\ \bibnamefont
  {Yao}}, \bibinfo {author} {\bibfnamefont {C.~R.}\ \bibnamefont {Laumann}},
  \bibinfo {author} {\bibfnamefont {S.}~\bibnamefont {Gopalakrishnan}},
  \bibinfo {author} {\bibfnamefont {M.}~\bibnamefont {Knap}}, \bibinfo {author}
  {\bibfnamefont {M.}~\bibnamefont {M{\"u}ller}}, \bibinfo {author}
  {\bibfnamefont {E.~A.}\ \bibnamefont {Demler}}, \ and\ \bibinfo {author}
  {\bibfnamefont {M.~D.}\ \bibnamefont {Lukin}},\ }\href {\doibase
  10.1103/PhysRevLett.113.243002} {\bibfield  {journal} {\bibinfo  {journal}
  {Phys. Rev. Lett.}\ }\textbf {\bibinfo {volume} {113}},\ \bibinfo {pages}
  {243002} (\bibinfo {year} {2014})}\BibitemShut {NoStop}%
\bibitem [{\citenamefont {Hauke}\ and\ \citenamefont {Heyl}(2015)}]{Hauke2014}%
  \BibitemOpen
  \bibfield  {author} {\bibinfo {author} {\bibfnamefont {P.}~\bibnamefont
  {Hauke}}\ and\ \bibinfo {author} {\bibfnamefont {M.}~\bibnamefont {Heyl}},\
  }\href@noop {} {\bibfield  {journal} {\bibinfo  {journal} {Phys. Rev. B}\
  }\textbf {\bibinfo {volume} {92}},\ \bibinfo {pages} {134204} (\bibinfo
  {year} {2015})}\BibitemShut {NoStop}%
\bibitem [{\citenamefont {Burin}(2015)}]{Burin2015}%
  \BibitemOpen
  \bibfield  {author} {\bibinfo {author} {\bibfnamefont {A.~L.}\ \bibnamefont
  {Burin}},\ }\href {\doibase 10.1103/PhysRevB.92.104428} {\bibfield  {journal}
  {\bibinfo  {journal} {Phys. Rev. B}\ }\textbf {\bibinfo {volume} {92}},\
  \bibinfo {pages} {104428} (\bibinfo {year} {2015})}\BibitemShut {NoStop}%
\bibitem [{\citenamefont {{Smith}}\ \emph {et~al.}(2016)\citenamefont
  {{Smith}}, \citenamefont {{Lee}}, \citenamefont {{Richerme}}, \citenamefont
  {{Neyenhuis}}, \citenamefont {{Hess}}, \citenamefont {{Hauke}}, \citenamefont
  {{Heyl}}, \citenamefont {{Huse}},\ and\ \citenamefont
  {{Monroe}}}]{Smith2015}%
  \BibitemOpen
  \bibfield  {author} {\bibinfo {author} {\bibfnamefont {J.}~\bibnamefont
  {{Smith}}}, \bibinfo {author} {\bibfnamefont {A.}~\bibnamefont {{Lee}}},
  \bibinfo {author} {\bibfnamefont {P.}~\bibnamefont {{Richerme}}}, \bibinfo
  {author} {\bibfnamefont {B.}~\bibnamefont {{Neyenhuis}}}, \bibinfo {author}
  {\bibfnamefont {P.~W.}\ \bibnamefont {{Hess}}}, \bibinfo {author}
  {\bibfnamefont {P.}~\bibnamefont {{Hauke}}}, \bibinfo {author} {\bibfnamefont
  {M.}~\bibnamefont {{Heyl}}}, \bibinfo {author} {\bibfnamefont {D.~A.}\
  \bibnamefont {{Huse}}}, \ and\ \bibinfo {author} {\bibfnamefont
  {C.}~\bibnamefont {{Monroe}}},\ }\href {\doibase 10.1038/nphys3783}
  {\bibfield  {journal} {\bibinfo  {journal} {Nature Physics}\ } (\bibinfo
  {year} {2016}),\ 10.1038/nphys3783}\BibitemShut {NoStop}%
\bibitem [{\citenamefont {Gogolin}\ \emph {et~al.}(2011)\citenamefont
  {Gogolin}, \citenamefont {M\"uller},\ and\ \citenamefont
  {Eisert}}]{Gogolin2011}%
  \BibitemOpen
  \bibfield  {author} {\bibinfo {author} {\bibfnamefont {C.}~\bibnamefont
  {Gogolin}}, \bibinfo {author} {\bibfnamefont {M.~P.}\ \bibnamefont
  {M\"uller}}, \ and\ \bibinfo {author} {\bibfnamefont {J.}~\bibnamefont
  {Eisert}},\ }\href {\doibase 10.1103/PhysRevLett.106.040401} {\bibfield
  {journal} {\bibinfo  {journal} {Phys. Rev. Lett.}\ }\textbf {\bibinfo
  {volume} {106}},\ \bibinfo {pages} {040401} (\bibinfo {year}
  {2011})}\BibitemShut {NoStop}%
\bibitem [{\citenamefont {Schreiber}\ \emph {et~al.}(2015)\citenamefont
  {Schreiber}, \citenamefont {Hodgman}, \citenamefont {Bordia}, \citenamefont
  {L{\"u}schen}, \citenamefont {Fischer}, \citenamefont {Vosk}, \citenamefont
  {Altman}, \citenamefont {Schneider},\ and\ \citenamefont
  {Bloch}}]{Schreiber2015}%
  \BibitemOpen
  \bibfield  {author} {\bibinfo {author} {\bibfnamefont {M.}~\bibnamefont
  {Schreiber}}, \bibinfo {author} {\bibfnamefont {S.~S.}\ \bibnamefont
  {Hodgman}}, \bibinfo {author} {\bibfnamefont {P.}~\bibnamefont {Bordia}},
  \bibinfo {author} {\bibfnamefont {H.~P.}\ \bibnamefont {L{\"u}schen}},
  \bibinfo {author} {\bibfnamefont {M.~H.}\ \bibnamefont {Fischer}}, \bibinfo
  {author} {\bibfnamefont {R.}~\bibnamefont {Vosk}}, \bibinfo {author}
  {\bibfnamefont {E.}~\bibnamefont {Altman}}, \bibinfo {author} {\bibfnamefont
  {U.}~\bibnamefont {Schneider}}, \ and\ \bibinfo {author} {\bibfnamefont
  {I.}~\bibnamefont {Bloch}},\ }\href {\doibase 10.1126/science.aaa7432}
  {\bibfield  {journal} {\bibinfo  {journal} {Science}\ }\textbf {\bibinfo
  {volume} {349}},\ \bibinfo {pages} {842} (\bibinfo {year}
  {2015})}\BibitemShut {NoStop}%
\bibitem [{\citenamefont {Nogrette}\ \emph {et~al.}(2014)\citenamefont
  {Nogrette}, \citenamefont {Labuhn}, \citenamefont {Ravets}, \citenamefont
  {Barredo}, \citenamefont {B\'eguin}, \citenamefont {Vernier}, \citenamefont
  {Lahaye},\ and\ \citenamefont {Browaeys}}]{Nogrette2014}%
  \BibitemOpen
  \bibfield  {author} {\bibinfo {author} {\bibfnamefont {F.}~\bibnamefont
  {Nogrette}}, \bibinfo {author} {\bibfnamefont {H.}~\bibnamefont {Labuhn}},
  \bibinfo {author} {\bibfnamefont {S.}~\bibnamefont {Ravets}}, \bibinfo
  {author} {\bibfnamefont {D.}~\bibnamefont {Barredo}}, \bibinfo {author}
  {\bibfnamefont {L.}~\bibnamefont {B\'eguin}}, \bibinfo {author}
  {\bibfnamefont {A.}~\bibnamefont {Vernier}}, \bibinfo {author} {\bibfnamefont
  {T.}~\bibnamefont {Lahaye}}, \ and\ \bibinfo {author} {\bibfnamefont
  {A.}~\bibnamefont {Browaeys}},\ }\href {\doibase 10.1103/PhysRevX.4.021034}
  {\bibfield  {journal} {\bibinfo  {journal} {Phys. Rev. X}\ }\textbf {\bibinfo
  {volume} {4}},\ \bibinfo {pages} {021034} (\bibinfo {year}
  {2014})}\BibitemShut {NoStop}%
\bibitem [{\citenamefont {Labuhn}\ \emph {et~al.}(2014)\citenamefont {Labuhn},
  \citenamefont {Ravets}, \citenamefont {Barredo}, \citenamefont {B\'eguin},
  \citenamefont {Nogrette}, \citenamefont {Lahaye},\ and\ \citenamefont
  {Browaeys}}]{Labuhn2014}%
  \BibitemOpen
  \bibfield  {author} {\bibinfo {author} {\bibfnamefont {H.}~\bibnamefont
  {Labuhn}}, \bibinfo {author} {\bibfnamefont {S.}~\bibnamefont {Ravets}},
  \bibinfo {author} {\bibfnamefont {D.}~\bibnamefont {Barredo}}, \bibinfo
  {author} {\bibfnamefont {L.}~\bibnamefont {B\'eguin}}, \bibinfo {author}
  {\bibfnamefont {F.}~\bibnamefont {Nogrette}}, \bibinfo {author}
  {\bibfnamefont {T.}~\bibnamefont {Lahaye}}, \ and\ \bibinfo {author}
  {\bibfnamefont {A.}~\bibnamefont {Browaeys}},\ }\href {\doibase
  10.1103/PhysRevA.90.023415} {\bibfield  {journal} {\bibinfo  {journal} {Phys.
  Rev. A}\ }\textbf {\bibinfo {volume} {90}},\ \bibinfo {pages} {023415}
  (\bibinfo {year} {2014})}\BibitemShut {NoStop}%
\bibitem [{\citenamefont {Barredo}\ \emph {et~al.}(2015)\citenamefont
  {Barredo}, \citenamefont {Labuhn}, \citenamefont {Ravets}, \citenamefont
  {Lahaye}, \citenamefont {Browaeys},\ and\ \citenamefont
  {Adams}}]{Barredo2015}%
  \BibitemOpen
  \bibfield  {author} {\bibinfo {author} {\bibfnamefont {D.}~\bibnamefont
  {Barredo}}, \bibinfo {author} {\bibfnamefont {H.}~\bibnamefont {Labuhn}},
  \bibinfo {author} {\bibfnamefont {S.}~\bibnamefont {Ravets}}, \bibinfo
  {author} {\bibfnamefont {T.}~\bibnamefont {Lahaye}}, \bibinfo {author}
  {\bibfnamefont {A.}~\bibnamefont {Browaeys}}, \ and\ \bibinfo {author}
  {\bibfnamefont {C.~S.}\ \bibnamefont {Adams}},\ }\href {\doibase
  10.1103/PhysRevLett.114.113002} {\bibfield  {journal} {\bibinfo  {journal}
  {Phys. Rev. Lett.}\ }\textbf {\bibinfo {volume} {114}},\ \bibinfo {pages}
  {113002} (\bibinfo {year} {2015})}\BibitemShut {NoStop}%
\bibitem [{\citenamefont {{Labuhn}}\ \emph {et~al.}(2016)\citenamefont
  {{Labuhn}}, \citenamefont {{Barredo}}, \citenamefont {{Ravets}},
  \citenamefont {{de L{\'e}s{\'e}leuc}}, \citenamefont {{Macr{\`i}}},
  \citenamefont {{Lahaye}},\ and\ \citenamefont {{Browaeys}}}]{Labuhn2015}%
  \BibitemOpen
  \bibfield  {author} {\bibinfo {author} {\bibfnamefont {H.}~\bibnamefont
  {{Labuhn}}}, \bibinfo {author} {\bibfnamefont {D.}~\bibnamefont {{Barredo}}},
  \bibinfo {author} {\bibfnamefont {S.}~\bibnamefont {{Ravets}}}, \bibinfo
  {author} {\bibfnamefont {S.}~\bibnamefont {{de L{\'e}s{\'e}leuc}}}, \bibinfo
  {author} {\bibfnamefont {T.}~\bibnamefont {{Macr{\`i}}}}, \bibinfo {author}
  {\bibfnamefont {T.}~\bibnamefont {{Lahaye}}}, \ and\ \bibinfo {author}
  {\bibfnamefont {A.}~\bibnamefont {{Browaeys}}},\ }\href {\doibase
  10.1038/nature18274} {\bibfield  {journal} {\bibinfo  {journal} {Nature}\
  }\textbf {\bibinfo {volume} {534}},\ \bibinfo {pages} {667} (\bibinfo {year}
  {2016})}\BibitemShut {NoStop}%
\bibitem [{\citenamefont {{L\"ow}}\ \emph {et~al.}(2012)\citenamefont
  {{L\"ow}}, \citenamefont {Weimer}, \citenamefont {Nipper}, \citenamefont
  {Balewski}, \citenamefont {Butscher}, \citenamefont {{B\"uchler}},\ and\
  \citenamefont {Pfau}}]{Rydberg2}%
  \BibitemOpen
  \bibfield  {author} {\bibinfo {author} {\bibfnamefont {R.}~\bibnamefont
  {{L\"ow}}}, \bibinfo {author} {\bibfnamefont {H.}~\bibnamefont {Weimer}},
  \bibinfo {author} {\bibfnamefont {J.}~\bibnamefont {Nipper}}, \bibinfo
  {author} {\bibfnamefont {J.~B.}\ \bibnamefont {Balewski}}, \bibinfo {author}
  {\bibfnamefont {B.}~\bibnamefont {Butscher}}, \bibinfo {author}
  {\bibfnamefont {H.~P.}\ \bibnamefont {{B\"uchler}}}, \ and\ \bibinfo {author}
  {\bibfnamefont {T.}~\bibnamefont {Pfau}},\ }\href@noop {} {\bibfield
  {journal} {\bibinfo  {journal} {J. Phys. B: At. Mol. Opt. Phys.}\ }\textbf
  {\bibinfo {volume} {45}},\ \bibinfo {pages} {113001} (\bibinfo {year}
  {2012})}\BibitemShut {NoStop}%
\bibitem [{\citenamefont {B\'eguin}\ \emph {et~al.}(2013)\citenamefont
  {B\'eguin}, \citenamefont {Vernier}, \citenamefont {Chicireanu},
  \citenamefont {Lahaye},\ and\ \citenamefont {Browaeys}}]{Beguin2013}%
  \BibitemOpen
  \bibfield  {author} {\bibinfo {author} {\bibfnamefont {L.}~\bibnamefont
  {B\'eguin}}, \bibinfo {author} {\bibfnamefont {A.}~\bibnamefont {Vernier}},
  \bibinfo {author} {\bibfnamefont {R.}~\bibnamefont {Chicireanu}}, \bibinfo
  {author} {\bibfnamefont {T.}~\bibnamefont {Lahaye}}, \ and\ \bibinfo {author}
  {\bibfnamefont {A.}~\bibnamefont {Browaeys}},\ }\href {\doibase
  10.1103/PhysRevLett.110.263201} {\bibfield  {journal} {\bibinfo  {journal}
  {Phys. Rev. Lett.}\ }\textbf {\bibinfo {volume} {110}},\ \bibinfo {pages}
  {263201} (\bibinfo {year} {2013})}\BibitemShut {NoStop}%
\bibitem [{\citenamefont {Ates}\ \emph {et~al.}(2007)\citenamefont {Ates},
  \citenamefont {Pohl}, \citenamefont {Pattard},\ and\ \citenamefont
  {Rost}}]{Ates07}%
  \BibitemOpen
  \bibfield  {author} {\bibinfo {author} {\bibfnamefont {C.}~\bibnamefont
  {Ates}}, \bibinfo {author} {\bibfnamefont {T.}~\bibnamefont {Pohl}}, \bibinfo
  {author} {\bibfnamefont {T.}~\bibnamefont {Pattard}}, \ and\ \bibinfo
  {author} {\bibfnamefont {J.~M.}\ \bibnamefont {Rost}},\ }\href {\doibase
  10.1103/PhysRevLett.98.023002} {\bibfield  {journal} {\bibinfo  {journal}
  {Phys. Rev. Lett.}\ }\textbf {\bibinfo {volume} {98}},\ \bibinfo {pages}
  {023002} (\bibinfo {year} {2007})}\BibitemShut {NoStop}%
\bibitem [{\citenamefont {Amthor}\ \emph {et~al.}(2010)\citenamefont {Amthor},
  \citenamefont {Giese}, \citenamefont {Hofmann},\ and\ \citenamefont
  {Weidem\"uller}}]{Amthor2010}%
  \BibitemOpen
  \bibfield  {author} {\bibinfo {author} {\bibfnamefont {T.}~\bibnamefont
  {Amthor}}, \bibinfo {author} {\bibfnamefont {C.}~\bibnamefont {Giese}},
  \bibinfo {author} {\bibfnamefont {C.~S.}\ \bibnamefont {Hofmann}}, \ and\
  \bibinfo {author} {\bibfnamefont {M.}~\bibnamefont {Weidem\"uller}},\ }\href
  {\doibase 10.1103/PhysRevLett.104.013001} {\bibfield  {journal} {\bibinfo
  {journal} {Phys. Rev. Lett.}\ }\textbf {\bibinfo {volume} {104}},\ \bibinfo
  {pages} {013001} (\bibinfo {year} {2010})}\BibitemShut {NoStop}%
\bibitem [{\citenamefont {Lesanovsky}\ and\ \citenamefont
  {Garrahan}(2014)}]{Lesanovsky14}%
  \BibitemOpen
  \bibfield  {author} {\bibinfo {author} {\bibfnamefont {I.}~\bibnamefont
  {Lesanovsky}}\ and\ \bibinfo {author} {\bibfnamefont {J.~P.}\ \bibnamefont
  {Garrahan}},\ }\href@noop {} {\bibfield  {journal} {\bibinfo  {journal}
  {Phys. Rev. A}\ }\textbf {\bibinfo {volume} {90}},\ \bibinfo {pages} {011603}
  (\bibinfo {year} {2014})}\BibitemShut {NoStop}%
\bibitem [{\citenamefont {Lesanovsky}\ and\ \citenamefont
  {Garrahan}(2013)}]{PRL-KinC}%
  \BibitemOpen
  \bibfield  {author} {\bibinfo {author} {\bibfnamefont {I.}~\bibnamefont
  {Lesanovsky}}\ and\ \bibinfo {author} {\bibfnamefont {J.~P.}\ \bibnamefont
  {Garrahan}},\ }\href@noop {} {\bibfield  {journal} {\bibinfo  {journal}
  {Phys. Rev. Lett.}\ }\textbf {\bibinfo {volume} {111}},\ \bibinfo {pages}
  {215305} (\bibinfo {year} {2013})}\BibitemShut {NoStop}%
\bibitem [{\citenamefont {Valado}\ \emph {et~al.}(2016)\citenamefont {Valado},
  \citenamefont {Simonelli}, \citenamefont {Hoogerland}, \citenamefont
  {Lesanovsky}, \citenamefont {Garrahan}, \citenamefont {Arimondo},
  \citenamefont {Ciampini},\ and\ \citenamefont {Morsch}}]{Valado2015}%
  \BibitemOpen
  \bibfield  {author} {\bibinfo {author} {\bibfnamefont {M.~M.}\ \bibnamefont
  {Valado}}, \bibinfo {author} {\bibfnamefont {C.}~\bibnamefont {Simonelli}},
  \bibinfo {author} {\bibfnamefont {M.~D.}\ \bibnamefont {Hoogerland}},
  \bibinfo {author} {\bibfnamefont {I.}~\bibnamefont {Lesanovsky}}, \bibinfo
  {author} {\bibfnamefont {J.~P.}\ \bibnamefont {Garrahan}}, \bibinfo {author}
  {\bibfnamefont {E.}~\bibnamefont {Arimondo}}, \bibinfo {author}
  {\bibfnamefont {D.}~\bibnamefont {Ciampini}}, \ and\ \bibinfo {author}
  {\bibfnamefont {O.}~\bibnamefont {Morsch}},\ }\href {\doibase
  10.1103/PhysRevA.93.040701} {\bibfield  {journal} {\bibinfo  {journal} {Phys.
  Rev. A}\ }\textbf {\bibinfo {volume} {93}},\ \bibinfo {pages} {040701}
  (\bibinfo {year} {2016})}\BibitemShut {NoStop}%
\bibitem [{\citenamefont {Izrailev}\ \emph {et~al.}(1995)\citenamefont
  {Izrailev}, \citenamefont {Kottos},\ and\ \citenamefont
  {Tsironis}}]{Izrailev1995}%
  \BibitemOpen
  \bibfield  {author} {\bibinfo {author} {\bibfnamefont {F.~M.}\ \bibnamefont
  {Izrailev}}, \bibinfo {author} {\bibfnamefont {T.}~\bibnamefont {Kottos}}, \
  and\ \bibinfo {author} {\bibfnamefont {G.~P.}\ \bibnamefont {Tsironis}},\
  }\href {\doibase 10.1103/PhysRevB.52.3274} {\bibfield  {journal} {\bibinfo
  {journal} {Phys. Rev. B}\ }\textbf {\bibinfo {volume} {52}},\ \bibinfo
  {pages} {3274} (\bibinfo {year} {1995})}\BibitemShut {NoStop}%
\bibitem [{\citenamefont {Izrailev}\ and\ \citenamefont
  {Krokhin}(1999)}]{Izrailev1999}%
  \BibitemOpen
  \bibfield  {author} {\bibinfo {author} {\bibfnamefont {F.~M.}\ \bibnamefont
  {Izrailev}}\ and\ \bibinfo {author} {\bibfnamefont {A.~A.}\ \bibnamefont
  {Krokhin}},\ }\href {\doibase 10.1103/PhysRevLett.82.4062} {\bibfield
  {journal} {\bibinfo  {journal} {Phys. Rev. Lett.}\ }\textbf {\bibinfo
  {volume} {82}},\ \bibinfo {pages} {4062} (\bibinfo {year}
  {1999})}\BibitemShut {NoStop}%
\bibitem [{\citenamefont {Furstenberg}\ and\ \citenamefont
  {Kesten}(1960)}]{Furstenberg1960}%
  \BibitemOpen
  \bibfield  {author} {\bibinfo {author} {\bibfnamefont {H.}~\bibnamefont
  {Furstenberg}}\ and\ \bibinfo {author} {\bibfnamefont {H.}~\bibnamefont
  {Kesten}},\ }\href {\doibase 10.1214/aoms/1177705909} {\bibfield  {journal}
  {\bibinfo  {journal} {Ann. Math. Statist.}\ }\textbf {\bibinfo {volume}
  {31}},\ \bibinfo {pages} {457} (\bibinfo {year} {1960})}\BibitemShut
  {NoStop}%
\bibitem [{\citenamefont {Flores}(1989)}]{Flores1989}%
  \BibitemOpen
  \bibfield  {author} {\bibinfo {author} {\bibfnamefont {J.~C.}\ \bibnamefont
  {Flores}},\ }\href {http://stacks.iop.org/0953-8984/1/i=44/a=017} {\bibfield
  {journal} {\bibinfo  {journal} {Journal of Physics: Condensed Matter}\
  }\textbf {\bibinfo {volume} {1}},\ \bibinfo {pages} {8471} (\bibinfo {year}
  {1989})}\BibitemShut {NoStop}%
\bibitem [{\citenamefont {Dunlap}\ \emph {et~al.}(1990)\citenamefont {Dunlap},
  \citenamefont {Wu},\ and\ \citenamefont {Phillips}}]{Dunlap1990}%
  \BibitemOpen
  \bibfield  {author} {\bibinfo {author} {\bibfnamefont {D.~H.}\ \bibnamefont
  {Dunlap}}, \bibinfo {author} {\bibfnamefont {H.-L.}\ \bibnamefont {Wu}}, \
  and\ \bibinfo {author} {\bibfnamefont {P.~W.}\ \bibnamefont {Phillips}},\
  }\href {\doibase 10.1103/PhysRevLett.65.88} {\bibfield  {journal} {\bibinfo
  {journal} {Phys. Rev. Lett.}\ }\textbf {\bibinfo {volume} {65}},\ \bibinfo
  {pages} {88} (\bibinfo {year} {1990})}\BibitemShut {NoStop}%
\bibitem [{\citenamefont {Bovier}(1992)}]{Bovier1992}%
  \BibitemOpen
  \bibfield  {author} {\bibinfo {author} {\bibfnamefont {A.}~\bibnamefont
  {Bovier}},\ }\href {http://stacks.iop.org/0305-4470/25/i=5/a=011} {\bibfield
  {journal} {\bibinfo  {journal} {Journal of Physics A: Mathematical and
  General}\ }\textbf {\bibinfo {volume} {25}},\ \bibinfo {pages} {1021}
  (\bibinfo {year} {1992})}\BibitemShut {NoStop}%
\bibitem [{\citenamefont {De~Bi{\`e}vre}\ and\ \citenamefont
  {Germinet}(2000)}]{DeBievre2000}%
  \BibitemOpen
  \bibfield  {author} {\bibinfo {author} {\bibfnamefont {S.}~\bibnamefont
  {De~Bi{\`e}vre}}\ and\ \bibinfo {author} {\bibfnamefont {F.}~\bibnamefont
  {Germinet}},\ }\href {\doibase 10.1023/A:1018615728507} {\bibfield  {journal}
  {\bibinfo  {journal} {Journal of Statistical Physics}\ }\textbf {\bibinfo
  {volume} {98}},\ \bibinfo {pages} {1135} (\bibinfo {year}
  {2000})}\BibitemShut {NoStop}%
\bibitem [{\citenamefont {Mattioli}\ \emph {et~al.}(2015)\citenamefont
  {Mattioli}, \citenamefont {Gl{\"a}tzle},\ and\ \citenamefont
  {Lechner}}]{Mattioli2015}%
  \BibitemOpen
  \bibfield  {author} {\bibinfo {author} {\bibfnamefont {M.}~\bibnamefont
  {Mattioli}}, \bibinfo {author} {\bibfnamefont {A.~W.}\ \bibnamefont
  {Gl{\"a}tzle}}, \ and\ \bibinfo {author} {\bibfnamefont {W.}~\bibnamefont
  {Lechner}},\ }\href {http://stacks.iop.org/1367-2630/17/i=11/a=113039}
  {\bibfield  {journal} {\bibinfo  {journal} {New Journal of Physics}\ }\textbf
  {\bibinfo {volume} {17}},\ \bibinfo {pages} {113039} (\bibinfo {year}
  {2015})}\BibitemShut {NoStop}%
\bibitem [{\citenamefont {Boada}\ \emph {et~al.}(2012)\citenamefont {Boada},
  \citenamefont {Celi}, \citenamefont {Latorre},\ and\ \citenamefont
  {Lewenstein}}]{Boada:2012}%
  \BibitemOpen
  \bibfield  {author} {\bibinfo {author} {\bibfnamefont {O.}~\bibnamefont
  {Boada}}, \bibinfo {author} {\bibfnamefont {A.}~\bibnamefont {Celi}},
  \bibinfo {author} {\bibfnamefont {J.~I.}\ \bibnamefont {Latorre}}, \ and\
  \bibinfo {author} {\bibfnamefont {M.}~\bibnamefont {Lewenstein}},\ }\href
  {\doibase 10.1103/PhysRevLett.108.133001} {\bibfield  {journal} {\bibinfo
  {journal} {Phys. Rev. Lett.}\ }\textbf {\bibinfo {volume} {108}},\ \bibinfo
  {pages} {133001} (\bibinfo {year} {2012})}\BibitemShut {NoStop}%
\bibitem [{\citenamefont {Celi}\ \emph {et~al.}(2014)\citenamefont {Celi},
  \citenamefont {Massignan}, \citenamefont {Ruseckas}, \citenamefont {Goldman},
  \citenamefont {Spielman}, \citenamefont {Juzeli\ifmmode~\bar{u}\else
  \={u}\fi{}nas},\ and\ \citenamefont {Lewenstein}}]{Celi:2014}%
  \BibitemOpen
  \bibfield  {author} {\bibinfo {author} {\bibfnamefont {A.}~\bibnamefont
  {Celi}}, \bibinfo {author} {\bibfnamefont {P.}~\bibnamefont {Massignan}},
  \bibinfo {author} {\bibfnamefont {J.}~\bibnamefont {Ruseckas}}, \bibinfo
  {author} {\bibfnamefont {N.}~\bibnamefont {Goldman}}, \bibinfo {author}
  {\bibfnamefont {I.~B.}\ \bibnamefont {Spielman}}, \bibinfo {author}
  {\bibfnamefont {G.}~\bibnamefont {Juzeli\ifmmode~\bar{u}\else
  \={u}\fi{}nas}}, \ and\ \bibinfo {author} {\bibfnamefont {M.}~\bibnamefont
  {Lewenstein}},\ }\href {\doibase 10.1103/PhysRevLett.112.043001} {\bibfield
  {journal} {\bibinfo  {journal} {Phys. Rev. Lett.}\ }\textbf {\bibinfo
  {volume} {112}},\ \bibinfo {pages} {043001} (\bibinfo {year}
  {2014})}\BibitemShut {NoStop}%
\bibitem [{\citenamefont {Price}\ \emph {et~al.}(2015)\citenamefont {Price},
  \citenamefont {Zilberberg}, \citenamefont {Ozawa}, \citenamefont
  {Carusotto},\ and\ \citenamefont {Goldman}}]{Price:2015}%
  \BibitemOpen
  \bibfield  {author} {\bibinfo {author} {\bibfnamefont {H.~M.}\ \bibnamefont
  {Price}}, \bibinfo {author} {\bibfnamefont {O.}~\bibnamefont {Zilberberg}},
  \bibinfo {author} {\bibfnamefont {T.}~\bibnamefont {Ozawa}}, \bibinfo
  {author} {\bibfnamefont {I.}~\bibnamefont {Carusotto}}, \ and\ \bibinfo
  {author} {\bibfnamefont {N.}~\bibnamefont {Goldman}},\ }\href {\doibase
  10.1103/PhysRevLett.115.195303} {\bibfield  {journal} {\bibinfo  {journal}
  {Phys. Rev. Lett.}\ }\textbf {\bibinfo {volume} {115}},\ \bibinfo {pages}
  {195303} (\bibinfo {year} {2015})}\BibitemShut {NoStop}%
\bibitem [{\citenamefont {Mancini}\ \emph {et~al.}(2015)\citenamefont
  {Mancini}, \citenamefont {Pagano}, \citenamefont {Cappellini}, \citenamefont
  {Livi}, \citenamefont {Rider}, \citenamefont {Catani}, \citenamefont {Sias},
  \citenamefont {Zoller}, \citenamefont {Inguscio}, \citenamefont {Dalmonte}
  \emph {et~al.}}]{Mancini:2015}%
  \BibitemOpen
  \bibfield  {author} {\bibinfo {author} {\bibfnamefont {M.}~\bibnamefont
  {Mancini}}, \bibinfo {author} {\bibfnamefont {G.}~\bibnamefont {Pagano}},
  \bibinfo {author} {\bibfnamefont {G.}~\bibnamefont {Cappellini}}, \bibinfo
  {author} {\bibfnamefont {L.}~\bibnamefont {Livi}}, \bibinfo {author}
  {\bibfnamefont {M.}~\bibnamefont {Rider}}, \bibinfo {author} {\bibfnamefont
  {J.}~\bibnamefont {Catani}}, \bibinfo {author} {\bibfnamefont
  {C.}~\bibnamefont {Sias}}, \bibinfo {author} {\bibfnamefont {P.}~\bibnamefont
  {Zoller}}, \bibinfo {author} {\bibfnamefont {M.}~\bibnamefont {Inguscio}},
  \bibinfo {author} {\bibfnamefont {M.}~\bibnamefont {Dalmonte}},  \emph
  {et~al.},\ }\href@noop {} {\bibfield  {journal} {\bibinfo  {journal}
  {Science}\ }\textbf {\bibinfo {volume} {349}},\ \bibinfo {pages} {1510}
  (\bibinfo {year} {2015})}\BibitemShut {NoStop}%
\bibitem [{\citenamefont {Stuhl}\ \emph {et~al.}(2015)\citenamefont {Stuhl},
  \citenamefont {Lu}, \citenamefont {Aycock}, \citenamefont {Genkina},\ and\
  \citenamefont {Spielman}}]{Stuhl:2015}%
  \BibitemOpen
  \bibfield  {author} {\bibinfo {author} {\bibfnamefont {B.}~\bibnamefont
  {Stuhl}}, \bibinfo {author} {\bibfnamefont {H.-I.}\ \bibnamefont {Lu}},
  \bibinfo {author} {\bibfnamefont {L.}~\bibnamefont {Aycock}}, \bibinfo
  {author} {\bibfnamefont {D.}~\bibnamefont {Genkina}}, \ and\ \bibinfo
  {author} {\bibfnamefont {I.}~\bibnamefont {Spielman}},\ }\href@noop {}
  {\bibfield  {journal} {\bibinfo  {journal} {Science}\ }\textbf {\bibinfo
  {volume} {349}},\ \bibinfo {pages} {1514} (\bibinfo {year}
  {2015})}\BibitemShut {NoStop}%
\bibitem [{\citenamefont {Barbarino}\ \emph {et~al.}(2016)\citenamefont
  {Barbarino}, \citenamefont {Taddia}, \citenamefont {Rossini}, \citenamefont
  {Mazza},\ and\ \citenamefont {Fazio}}]{Barbarino:2016}%
  \BibitemOpen
  \bibfield  {author} {\bibinfo {author} {\bibfnamefont {S.}~\bibnamefont
  {Barbarino}}, \bibinfo {author} {\bibfnamefont {L.}~\bibnamefont {Taddia}},
  \bibinfo {author} {\bibfnamefont {D.}~\bibnamefont {Rossini}}, \bibinfo
  {author} {\bibfnamefont {L.}~\bibnamefont {Mazza}}, \ and\ \bibinfo {author}
  {\bibfnamefont {R.}~\bibnamefont {Fazio}},\ }\href@noop {} {\bibfield
  {journal} {\bibinfo  {journal} {New Journal of Physics}\ }\textbf {\bibinfo
  {volume} {18}},\ \bibinfo {pages} {035010} (\bibinfo {year}
  {2016})}\BibitemShut {NoStop}%
\bibitem [{\citenamefont {Bell}\ and\ \citenamefont {Dean}(1970)}]{Bell1970}%
  \BibitemOpen
  \bibfield  {author} {\bibinfo {author} {\bibfnamefont {R.~J.}\ \bibnamefont
  {Bell}}\ and\ \bibinfo {author} {\bibfnamefont {P.}~\bibnamefont {Dean}},\
  }\href {\doibase 10.1039/DF9705000055} {\bibfield  {journal} {\bibinfo
  {journal} {Discuss. Faraday Soc.}\ }\textbf {\bibinfo {volume} {50}},\
  \bibinfo {pages} {55} (\bibinfo {year} {1970})}\BibitemShut {NoStop}%
\bibitem [{\citenamefont {da~Fonseca}\ and\ \citenamefont
  {Petronilho}(2001)}]{MatInverse}%
  \BibitemOpen
  \bibfield  {author} {\bibinfo {author} {\bibfnamefont {C.}~\bibnamefont
  {da~Fonseca}}\ and\ \bibinfo {author} {\bibfnamefont {J.}~\bibnamefont
  {Petronilho}},\ }\href {\doibase
  http://dx.doi.org/10.1016/S0024-3795(00)00289-5} {\bibfield  {journal}
  {\bibinfo  {journal} {Linear Algebra and its Applications}\ }\textbf
  {\bibinfo {volume} {325}},\ \bibinfo {pages} {7 } (\bibinfo {year}
  {2001})}\BibitemShut {NoStop}%
\end{thebibliography}
